\documentclass[draft]{agujournal2019}
\usepackage{url} %this package should fix any errors with URLs in refs.
\usepackage{lineno}
\usepackage[inline]{trackchanges} %for better track changes. finalnew option will compile document with changes incorporated.
\usepackage{soul}
\usepackage{epstopdf}%This line makes .eps figures into .pdf - please comment out if not required.
\newcommand{\ef}[1]{{\color{black} #1}}
\newcommand{\efb}[1]{{\color{black} #1}}

\newcommand{\wa}[1]{{\color{black} #1}}
\newcommand{\wAssis}[1]{{\color{black} #1}}
\draftfalse

\journalname{JGR: Earth Surface}

\begin{document}

%% ------------------------------------------------------------------------ %%
%  Title
%
% (A title should be specific, informative, and brief. Use
% abbreviations only if they are defined in the abstract. Titles that
% start with general keywords then specific terms are optimized in
% searches)
%
%% ------------------------------------------------------------------------ %%

% Example: \title{This is a test title}

\title{Evolving dunes under flow reversals: from an initial heap toward an inverted dune}

%% ------------------------------------------------------------------------ %%
%
%  AUTHORS AND AFFILIATIONS
%
%% ------------------------------------------------------------------------ %%

% Authors are individuals who have significantly contributed to the
% research and preparation of the article. Group authors are allowed, if
% each author in the group is separately identified in an appendix.)

% List authors by first name or initial followed by last name and
% separated by commas. Use \affil{} to number affiliations, and
% \thanks{} for author notes.
% Additional author notes should be indicated with \thanks{} (for
% example, for current addresses).

% Example: \authors{A. B. Author\affil{1}\thanks{Current address, Antartica}, B. C. Author\affil{2,3}, and D. E.
% Author\affil{3,4}\thanks{Also funded by Monsanto.}}

\authors{W. R. Assis\affil{1,5}, E. M. Franklin\affil{1}, N. M. Vriend\affil{2,3,4,6}}

\affiliation{1}{School of Mechanical Engineering, UNICAMP - University of Campinas, Rua Mendeleyev, 200, Campinas, SP, Brazil}
\affiliation{2}{BP Institute, University of Cambridge, Cambridge, UK}
\affiliation{3}{Department of Applied Mathematics and Theoretical Physics, University of Cambridge, Cambridge, UK}
\affiliation{4}{Department of Earth Sciences, University of Cambridge, Cambridge, UK}
\affiliation{5}{Current address: Saint Anthony Falls Laboratory, University of Minnesota, Minneapolis, Minnesota, USA}
\affiliation{6}{Current address: Department of Mechanical Engineering, University of Colorado Boulder, Boulder, USA}
% \affiliation{2}{Second Affiliation}
% \affiliation{3}{Third Affiliation}
% \affiliation{4}{Fourth Affiliation}

%\affiliation{=number=}{=Affiliation Address=}
%(repeat as many times as is necessary)

%% Corresponding Author:
% Corresponding author mailing address and e-mail address:

% (include name and email addresses of the corresponding author.  More
% than one corresponding author is allowed in this LaTeX file and for
% publication; but only one corresponding author is allowed in our
% editorial system.)

% Example: \correspondingauthor{First and Last Name}{email@address.edu}

\correspondingauthor{Willian Assis}{righiassis@gmail.com}

%% Keypoints, final entry on title page.

%  List up to three key points (at least one is required)
%  Key Points summarize the main points and conclusions of the article
%  Each must be 140 characters or fewer with no special characters or punctuation and must be complete sentences

% Example:
% \begin{keypoints}
% \item	List up to three key points (at least one is required)
% \item	Key Points summarize the main points and conclusions of the article
% \item	Each must be 140 characters or fewer with no special characters or punctuation and must be complete sentences
% \end{keypoints}

\begin{keypoints}

\item Experiments show that 2D dunes grow and develop over a characteristic time that matches that of fully 3D barchan dunes

\item The morphodynamics of reversing dunes over time are revealed by fully reversing the flow direction and tracking the rebuilding and reshaping
\item Numerical simulations on a reversing 3D barchan show that its central slice behaves as the reversing 2D dunes

\end{keypoints}

%% ------------------------------------------------------------------------ %%
%
%  ABSTRACT and PLAIN LANGUAGE SUMMARY
%
% A good Abstract will begin with a short description of the problem
% being addressed, briefly describe the new data or analyses, then
% briefly states the main conclusion(s) and how they are supported and
% uncertainties.

% The Plain Language Summary should be written for a broad audience,
% including journalists and the science-interested public, that will not have 
% a background in your field.
%
% A Plain Language Summary is required in GRL, JGR: Planets, JGR: Biogeosciences,
% JGR: Oceans, G-Cubed, Reviews of Geophysics, and JAMES.
% see http://sharingscience.agu.org/creating-plain-language-summary/)
%
%% ------------------------------------------------------------------------ %%

%% \begin{abstract} starts the second page

\begin{abstract}
Sand dunes are ubiquitous in nature, and are found in abundance on Earth and other planetary environments. One of the most common types are crescent-shaped dunes known as barchans, whose mid-line could be assumed to behave as 2D dunes. In this work, we (i) compare the morphology of the mid-line of 3D barchans with 2D dunes; and (ii) track the evolution of 3D barchans and 2D dunes while reversing flow conditions. We performed experiments on 2D dunes in a 2D flume and Euler-Lagrange simulations of 3D bedforms. In all reversal experiments and simulations, the initial condition start with a conical heap deforming into a steady-state dune, which is then perturbed by reversing the flow, resulting in an inverted dune. We show that during the reversal the grains on the lee side immediately climb back onto the dune while its internal part and toe remain static, forming a new lee face of varying angle on the previous stoss slope. We show that (i) the characteristic time for the development of 2D dunes scales with that for 3D barchans, (ii) that the time for dune reversal is twice the time necessary to develop an initial triangular or conical heap to steady-state, and (iii) that a considerable part of grains remain static during the entire process. Our findings reveal the dynamics for dune reversal, and highlight that numerical computations of barchans based on 2D slices, which are more feasible in geophysical scales, predict realistic outcomes for the relevant time-scales.

\end{abstract}

\section*{Plain Language Summary}
Crescent-shaped dunes, known as barchans, are found in abundance on Earth and other planetary environments. Although their different shapes and manifestations intrigue us and produce fascinating images (such as in the latest images from Mars), the underlying physics still challenges us. Here we investigate two critical questions: (i) can we capture the relevant physical processes of 3D dunes in a 2D slice? and (ii) how does the dune morph over space and time upon flow reversal (e.g. the wind blowing from the opposite direction)? We research these questions by carrying out experiments with 2D dunes in a water flume and numerical simulations of 3D barchans, and flip-around our flow forcing to investigate flow reversal. We find that the typical development times for 2D and 3D dunes are equivalent and reveal details of the rebuilding processes of the dune upon flow reversals. Interestingly, the inversion time after flow reversal is twice that of the formation time of the initial heap, and a considerable part of grains remains static during the entire process. Our findings reveal the mechanisms for dune reversal and show that 2D simulations, which are simpler and faster, reproduce the underlying physics.

\pagebreak
%% ------------------------------------------------------------------------ %%
%
%  TEXT
%
%% ------------------------------------------------------------------------ %%

%%% Suggested section heads:
% \section{Introduction}
%
% The main text should start with an introduction. Except for short
% manuscripts (such as comments and replies), the text should be divided
% into sections, each with its own heading.

% Headings should be sentence fragments and do not begin with a
% lowercase letter or number. Examples of good headings are:

% \section{Materials and Methods}
% Here is text on Materials and Methods.
%
% \subsection{A descriptive heading about methods}
% More about Methods.
%
% \section{Data} (Or section title might be a descriptive heading about data)
%
% \section{Results} (Or section title might be a descriptive heading about the
% results)
%
% \section{Conclusions}

\section{Introduction}

Sand dunes are bedforms resulting from erosion and deposition of sand by the action of a fluid flow \cite{Bagnold_1, Hersen_1}, and they are frequently found on Earth, Mars and other celestial bodies \cite{Hersen_3, Elbelrhiti, Claudin_Andreotti, Parteli2, Courrech}. Among the most common are crescent-shaped dunes, known as barchans, that appear under a one-directional flow regime and when the quantity of available sand is limited i.e., they grow over a non-erodible ground over which the sand, once the dune is formed, does not cover the entire surface.

Given the abundance of barchans present in nature, a considerable number of field measurements, experiments, and numerical simulations were conducted over the last decades to better understand those bedforms \cite{Sauermann_1, Hersen_1, Andreotti_1, Hersen_2, Hersen_3, Kroy_B, Kroy_C, Parteli, Andreotti_4, Franklin_8, Pahtz_1, Kidanemariam, Guignier, Parteli3, Khosronejad}, but only very few of them were carried out at the grain scale \cite{Alvarez3, Alvarez4, Alvarez6, Alvarez7, Assis, Assis2}. Most analytical models and numerical simulations are based on information from field measurements on aeolian dunes, so that they solve the problem at the bedform scale only by considering that the grains move mainly in the longitudinal direction (typical of aeolian saltation). For example, the first numerical simulations considered the granular system as a continuum medium, some of them modeling 2D dunes \cite{Sauermann_4, Kroy_A, Kroy_C}, and some others modeling 3D dunes as vertical slices that behave as 2D dunes \cite{Herrmann_Sauermann, Hersen_3, Kroy_B}. The latter hypothesize that grains move in the longitudinal direction (the same direction of the fluid flow), while transverse diffusion transfers a small quantity of mass between adjoining slices \cite{Hersen_3}. More recent models consider the 3D perturbations of the fluid flow, and, thus, the transverse sediment transport \cite{Schwammle, Duran2, Parteli6, Parteli4}, but always in the aeolian case, meaning that the transport of grains occurs mainly in the longitudinal direction. In addition, in some cases a transverse diffusion of grains needs to be included for obtaining the correct crescent shape \cite<even though the transverse flow is present,>{Schwammle}. Therefore, the continuum-sliced models are, in principle, valid for aeolian barchans, which consist of a large number of grains that are entrained mainly in longitudinal direction, with small lateral motion due to reptation \cite{Andreotti_1, Hersen_3}. However, this is not necessarily the case of subaqueous barchans, where transverse sediment transport can become significant \cite<even if reptation is absent, the grains being directly entrained by the fluid by rolling and sliding,>{Andreotti_1}.

Some recent works showed that the transverse motion of grains is important for subaqueous barchans \cite{Alvarez3, Alvarez4}, indicating that the picture of a 3D dune as connected slices must be refined in the subaqueous case. For instance, \citeA{Alvarez3, Alvarez4} measured experimentally the displacement of individual grains migrating to horns as an initial pile was deformed into a barchan dune. They found that most of those grains come from upstream regions on the periphery of the dune, within angles forming 105$^{\circ}$ and 160$^{\circ}$ and 210$^{\circ}$ and 260$^{\circ}$ with respect to the flow direction (0$^{\circ}$ pointing downstream). Those results were later corroborated by numerical simulations at the grain scale using large eddy simulation coupled with discrete element method \cite<LES-DEM,>{Alvarez6, Alvarez7, Lima2}. In this picture, grains migrating to horns have considerable transverse displacements (of the order of the dune size), contradicting, for subaqueous barchans, the models based on connected slices. Note, however, that the results show that grains going to horns do not come from the dune centerline.

Based on discrete simulations using a cellular automaton model, by employing 2D slices interconnected by a diffusion process, \citeA{Zhang_D} found that the residence time of grains within a barchan dune (the typical time that a grain remains as part of the dune), in particular in the central slice, is relatively large, being of the order of many turnover times of the barchan (10 turnover times, if we consider all the grains forming the barchan). They showed that the residence time is given by the surface of the longitudinal central slice of the dune divided by the input sand flux, and that its large value occurs because of a cyclic process: grains on the stoss side tend to disperse toward the laterals \cite<as also shown by the experiments of>{Assis2}, but are returned to the central region after avalanching on the lee side due to the curvature of the barchan dune. On the whole, \citeA{Zhang_D} showed that transverse mixing in the central slice is restricted by this dispersion-concentration mechanism, and proposed that the central slice contains most of the information (and memory) of the barchan morphodynamics. This result is not, in principle, in contradiction with those of \citeA{Alvarez3, Alvarez4}, since the latter found that grains populating the horns (and afterward leaving the barchan) do not come from the central slice.

Because the interior (e.g, the central slice) of real dunes is not accessible in experiments, \citeA{bacik2020wake, bacik2021stability, bacik2021dynamics} carried out experiments with 2D dunes in a narrow Couette-type circular water flume. \citeA{bacik2020wake} investigated how 2D dunes interact with each other under a turbulent water flow, and found that the turbulent structures of the flow trigger a long range dune-dune repulsion (preventing dune-dune collisions). Later, \citeA{bacik2021stability} inquired into the stability of a pair of dunes and proposed a parameter space where dune-dune interactions either stabilize or destabilize the initial configuration, and \citeA{bacik2021dynamics} showed how the presence of obstacles change the dune morphodynamics. Later, \citeA{Assis, Assis2} investigated experimentally barchan-barchan interactions and \citeA{Assis3} barchans interacting with dune-size obstacles, but in their experiments the grains inside the dune were not accessible. Therefore, the findings for 2D dunes can be proven valid for barchan dunes, they would represent a large advance toward understanding barchan fields. 

Our aim is to investigate whether subaqueous 3D barchan dunes can be represented as connected slices, in essence as the 2D dune as introduced in \citeA{bacik2020wake, bacik2021stability, bacik2021dynamics}, or whether, instead, the transverse sediment transport radically changes the physical behavior and needs to be accounted for. In addition, we are investigating whether the underlying physical processes of dune reversal leading to an inverted dune can be captured as a solely 2D process mimicking the mid-line of a 3D barchan dune \cite<field evidences of reversing dunes were reported recently,>{Gao}. We carry out experiments in the 2D flume on heaps and reversing dunes and complement these experiments with numerical simulations at the grain-scale, which allows us to analyze the central slice of 3D dunes. In our numerical simulations, we apply the same forcing procedure as in our experiments: (1) pile formation, (2) development to a steady-state dune, (3) flow reversal, and (4) equilibrating to a steady-state (reversed) dune. As the grains climb back up the lee side during the flow reversal stage, the internal part and the toe of the dune remain static while a new lee face with varying angle and length is formed on the former stoss slope. In this manuscript, we identify characteristic times and scales of this reversal process, and the areas where the grains are remobilized in this re-morphing process. Our findings reveal the mechanisms for dune reversal and provide a validation between experimental data and numerical simulations.

\section{Experimental Setup}

\begin{figure}[h!]
	\begin{center}
		\includegraphics[width=.95\linewidth]{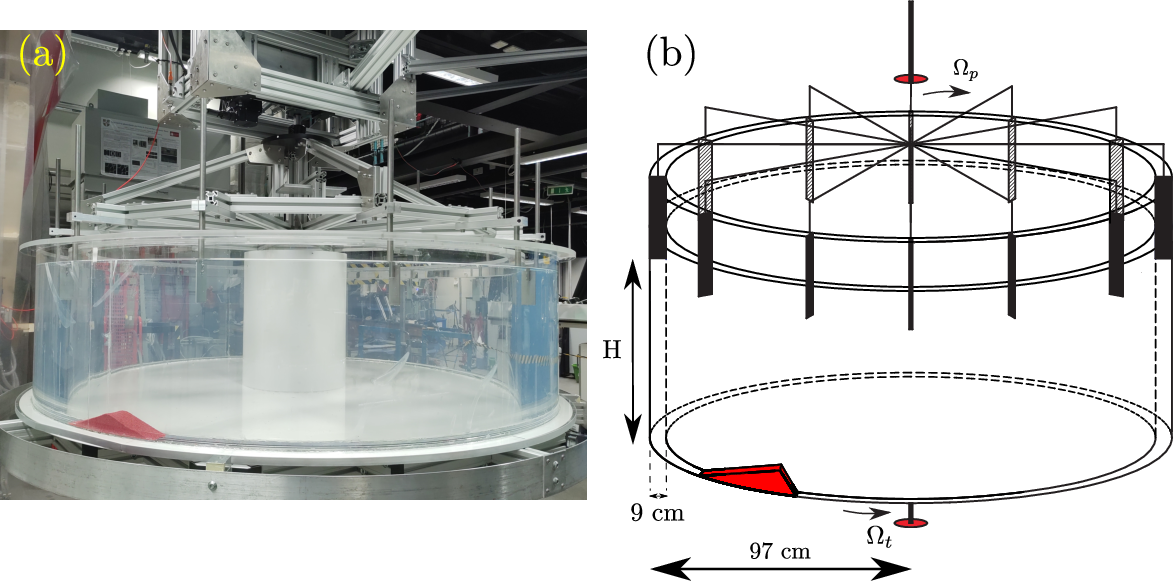}
	\end{center}
	\caption{ (a) Photograph and (b) Layout of the circular flume.}
	\label{fig:layout}
\end{figure}

The experimental setup is the same as used in \citeA{jarvis2022coarsening, bacik2020wake}, and consists of a periodic channel, a driving device, and an imaging system. The periodic channel is a circular flume with external and internal radii of 97 cm and 88 cm, respectively, filled to a level of 45.5 cm with water and particles, with parameters specified in the next paragraph. A rotating rig with 12 equidistant paddles submerged near the water surface is mounted above the flume, providing a shearing motion to the water, while the flume is connected to a counter-rotating turntable. Our tests begin by imposing a paddle rotation in the counter-clockwise direction (view from above) and a turntable rotation in the clockwise direction (we call this flow 0$^\circ$). After reaching a steady-state developed dune, we stop the flow, and after the system is at rest we revert both the paddle and turntable directions in order to obtain a reverse flow (180$^\circ$). Figures \ref{fig:layout}a and \ref{fig:layout}b show a photograph and the layout of the experimental setup, respectively. 

We used round glass spheres ($\rho_s$ = 2500 kg m$^{-3}$, approximately),
sieved to a diameter between 1.0 mm $\leq$ $d$  $\leq$ 1.3 mm, for which we consider the mean value as being $\bar d$ = 1.15 mm, and varied the total mass of the initial pile between 1 and 2 kg (see the supporting information for a photograph of the used particles). The flow direction was either 0$^\circ$ (initial flow) or 180$^\circ$ (reverse flow), and the mean water velocity varied within 0.81 m/s $\leq$ $U$ $\leq$  1.22 m/s. In here, the relative velocity between the table and paddles is $U$ = $R (\Omega_p-\Omega_t)$, the outer radius is $R$ = 97 cm (one entire revolution corresponds to a length of approximately 6 m), and $\Omega_p$ and $\Omega_t$ are the angular velocities of the paddles and table, respectively. The shear velocity $u_*$ is computed based on Equation 8 of \citeA{jarvis2022coarsening}, and was found to vary between 0.050 m s$^{-1}$ $\leq$ $u_*$ $\leq$ 0.103 m s$^{-1}$. The Reynolds number $Re$ = $Uw/\nu$ varied within $0.73$ $\times$ $10^5$ and $1.10$ $\times$ $10^5 $, where $w$ = 9 cm is the width of the channel and $\nu$ the kinematic viscosity of water. \ef{The inertial length $L_{drag}$ = $\left( \rho_p / \rho_f \right) d$, as proposed by \citeA{Hersen_1}, is approximately 2.5 mm (considering $d$ $\approx$ 1 mm), and the flux saturation length $L_{sat}$ $\approx$ $ 4.4 L_{drag}$ is 11 mm \cite{Andreotti_1, Andreotti_2, Claudin_Andreotti}. We note that, although the use of $L_{drag}$ for subaqueous dunes has been the object of debate \cite{Charru_3}, it is a simple expression that works reasonably well for comparing barchans in different environments, as shown by \citeA{Hersen_1}.} The paddle and water heights were fixed for all tests, being 34.5 and 45.5 cm, respectively. Table \ref{flume_conditions} summarizes the test conditions, and images from experiments are available in an open repository \cite{Supplemental2}. For a given velocity U, the exact angular velocities $\Omega_p$ and $\Omega_t$ were chosen empirically to reduce secondary flows in order to produce 2D dunes as symmetrical as possible (lateral-view images from 2D dunes are available in the supporting information).

\begin{table}[!ht]	
	\begin{center}
		\begin{tabular}{c c c c c c c c c c c}
			\hline\hline
			Case & Dune mass & $\Omega_p$ & $\Omega_t$ & $\Omega_p$ - $\Omega_t$ & $Re$ & Flow direction &Z  & L & $u_*$ & $t_c$ \\ 
			$\cdots$  & kg & rpm & rpm & rpm  & $\cdots$ & degrees & mm & mm & m s$^{-1}$ & s \\\hline
			\textit{a} & 2 & 4.60 & -3.40 & 8 &  $0.73$ $\times$ $10^5$ & 0& 70&383&0.050 & 180\\
			\textit{b} & 2 & 5.80 & -4.20 & 10 &  $0.91$ $\times$ $10^5$ & 0&72&407&0.077& 47\\
			\textit{c} & 2 & 7.00 & -5.00 & 12 & $1.10$ $\times$ $10^5$ & 0 &73&390&0.103&18\\
			\textit{d} & 1 & 4.60 & -3.40 & 8 & $0.73$ $\times$ $10^5$ & 0&49&264&0.050&120\\
			\textit{e} & 1 & 5.65 & -4.35 & 10 & $0.91$ $\times$ $10^5$ & 0&49&262&0.077&34\\
			\textit{f} & 1 & 6.85 & -5.15 & 12 & $1.10$ $\times$ $10^5$ & 0&51&278&0.103&13\\
			\textit{g} & 2 & -4.60 & 3.40 & -8 & $0.73$ $\times$ $10^5$ & 180&41&543&0.050&179\\
			\textit{h} & 2 & -5.80 & 4.20 & -10 & $0.91$ $\times$ $10^5$ & 180&59&499&0.077&47\\
			\textit{i} & 2 & -7.00 & 5.00 & -12 & $1.10$ $\times$ $10^5$ & 180&62&499&0.103&18\\
			\textit{j} & 1 & -4.60 & 3.40 & -8 & $0.73$ $\times$ $10^5$ & 180&32&363&0.050&117\\
			\textit{k} & 1 & -5.65 & 4.35 & -10 & $0.91$ $\times$ $10^5$ & 180&44&348&0.077&34\\
			\textit{l} & 1 & -6.85 & 5.15 & -12 & $1.10$ $\times$ $10^5$ & 180&48&338&0.103&13\\
		\end{tabular}
	\end{center}
	\caption{Label of tested cases, dune mass, angular velocity of paddles, angular velocity of the table, total angular velocity, channel Reynolds number $Re$,  flow orientation, initial height ($Z$), initial length ($L$), the shear velocity $u_*$, and the characteristic time $t_c$.}
	\label{flume_conditions}
\end{table}

A camera of complementary metal-oxide-semiconductor (CMOS) type was mounted on the ground (laboratoty frame of reference) with a lateral view (i.e., in the radial direction) of the flume. We used a ISVI black and white camera, capable of acquiring images at a maximum resolution of 12MP at 181 Hz (model IC-X12S-CXP), and a Nikon lens of 60 mm focal distance and F2.8 maximum aperture (model AF Micro Nikkor). In the experiments, we set the camera to operate with a region of interest (ROI) of 64 px $\times$ 1,024 px at a frequency of 200 Hz. The field of view was 6.6 mm $\times$ 105.5 mm, corresponding to a resolution of 9.7 px mm$^{-1}$. A column in the central axis of the rotating experiment (see Figure \ref{fig:layout}) was illuminated with lamps of light-emitting diode (LED), enabling a good contrast between the sediment layers and walls. The acquired images were afterward processed by numerical scripts that identify and reconstruct 2D profiles providing a dune shape (the used procedure needs only an overlap of images in the horizontal direction to assure that consecutive images are really correlated, avoiding the necessity of correcting optical aberration).

\section{Numerical Setup}

We carried out numerical simulations using CFD-DEM (computational fluid dynamics - discrete element method), in which we computed the formation of single barchans from initially conical piles and, after reaching a developed barchan shape, reversed the flow direction. Our simulations were performed at the grain scale by making use of LES (large eddy simulation) for CFD, which thus computed the mass (Equation \ref{mass}) and momentum (Equation \ref{mom}) equations for the fluid using meshes of the order of the grains' diameter,

\begin{equation}
\nabla\cdot\vec{u}_{f}=0 \, ,
\label{mass}
\end{equation}

\begin{equation}
\frac{\partial{\rho_{f}\vec{u}_{f}}}{\partial{t}} + \nabla \cdot (\rho_{f}\vec{u}_{f}\vec{u}_{f}) = -\nabla P + \nabla\cdot \vec{\vec{\tau}} + \rho_{f}\vec{g} - \vec{f}_{fp} \, ,
\label{mom}
\end{equation}

\noindent where $\vec{g}$ is the acceleration of gravity, $\vec{u}_{f}$ is the fluid velocity, $\rho_{f}$ is the fluid density, $P$ the fluid pressure, $\vec{\vec{\tau}}$ the deviatoric stress tensor of the fluid, and $\vec{f}_{fp}$ is the resultant of fluid forces acting on each grain by unit of fluid volume. The DEM solved the linear (Equation \ref{Fp}) and angular (Equation \ref{Tp}) momentum equations applied to each solid particle,

\begin{equation}
m_{p}\frac{d\vec{u}_{p}}{dt}= \vec{F}_{p}\, ,
\label{Fp}
\end{equation}

\begin{equation}
I_{p}\frac{d\vec{\omega}_{p}}{dt}=\vec{T}_{c}\, ,
\label{Tp}
\end{equation}

\noindent where, for each grain, $m_{p}$ is the mass, $\vec{u}_{p}$ is the velocity, $I_{p}$ is the moment of inertia, $\vec{\omega}_{p}$ is the angular velocity, $\vec{T}_{c}$ is the resultant of contact torques between solids, and $\vec{F}_{p}$ is the resultant force (weight, contact and fluid forces). We made use of the open-source code \mbox{CFDEM} \cite{Goniva} (www.cfdem.com), which couples the open-source CFD code OpenFOAM with the open-source DEM code LIGGGHTS \cite{Kloss, Berger}. A complete description of the fundamental and implemented equations, CFD meshes and convergence, DEM parameters, and tests can be found in \citeA{Lima2}.

The CFD domain is a 3D channel of size $L_x$ = 0.4 m, $L_y$ = $\delta$ = 0.025 m and $L_z$ = 0.1 m, where $x$, $y$ and $z$ are the longitudinal, vertical and spanwise directions, respectively, with periodic conditions in the longitudinal and spanwise directions. The vertical dimension of the domain, $L_y$ = $\delta$, corresponds to the channel half height (the real channel height being 2$\delta$), and the height of smallest meshes (close to the bottom boundary) was $\Delta y$ = 2.9 $\times$ 10$^{-4}$ m, which corresponds to $\Delta y / d$ = 1.46 (the values of $d$ used in the simulations are shown next). With the used mesh, the fluid flow close to the bed is resolved at a scale close to that of grains, capturing not only the recirculation that occurs downstream the dune crest, but also smaller vortices \cite{Lima2}. The fluid is water, flowing with a cross-sectional mean velocity $U$ = 0.28 m s$^{-1}$. The channel Reynolds number based on $U$, Re = $U 2\delta/\nu$, is 14,000, and the Reynolds number based on shear velocity $u_*$,  Re$_*$ = $u_* \delta/ \nu$, is 400, where $\nu$ is the kinematic viscosity (10$^{-6}$ m$^2$ s$^{-1}$ for water). The granular material consisted of  10$^5$ glass spheres randomly distributed, with sizes following a Gaussian distribution within 0.15 mm $\leq$ $d$ $\leq$ 0.25 mm. The coefficients of sliding friction $\mu$, rolling friction $\mu_r$ and restitution $e$, as well as the values of Poisson ratio $\sigma$, Young's modulus $E$ and density $\rho_p$ used in the simulations are shown in Table \ref{DEM_properties} \cite<extensive tests of these parameters are presented in>{Lima2}. We selected for the particles a solid wall boundary condition at the bottom boundary, and a free exit at the outlet. Note that no influx of grains was imposed, so that the bedform lose grains and decrease slightly in size along time. We note also that the numerical setup differs from the experimental one in terms of fluid flow, grain diameter, boundary conditions, and size of the system. While, on the one hand, to simulate barchans with a size comparable to the 2D experiments would be computationally unfeasible, on the other hand the numerical setup used has been extensively investigated and validated against experiments \cite{Lima2}. In addition, the use of periodic conditions for the grains (to be closer to the experimental setup) would imply that grains leaving the two horns would return and reach regions close to the flanks of the barchan dune, deforming it considerably. More details about the equations, parameters and meshes used in the simulations can be found in \citeA{Lima2}.

\begin{table}[!h]
	\centering
	\caption{Physical properties of DEM particles.}
	\begin{tabular}{lc}
		\hline
		\multicolumn{2}{l}{DEM properties} \\ \hline
	    Sliding Friction Coeff. $\mu$   & 0.6   \\
		Rolling Friction Coeff. $\mu_r$  & 0.00   \\
		Restitution Coef. $e$   & 0.1  \\
		Poisson Ratio $\sigma$   & 0.45   \\
		Young's Modulus $E$ (MPa) & 5    \\
		\wa{Density $\rho_p$ (kg m$^{-3}$)}  & 2500  \\ \hline
	\end{tabular}
	\label{DEM_properties}
\end{table}

The first step was to simulate a pure fluid (in the absence of solid particles) flowing in the periodic channel until reaching fully-developed turbulence, and store the output to be used as initial condition in the CFD-DEM  simulations (which are periodic only for the fluid). This step aimed at obtaining the initial conditions for the fluid with relatively low computational cost. Then, prior to each simulation, the grains are allowed to fall freely in stationary water, forming a conical heap in the channel center. Finally, the CFD-DEM simulations begin by imposing a turbulent water flow (whose initial condition was the previously stored fully-turbulent flow), which deforms the conical pile into a barchan dune. When a developed barchan is achieved, the flow is stopped and its direction reversed. Files with the setups used in our CFD-DEM simulations are available in an open repository \cite{Supplemental2}. We note that we carried out 3D instead of 2D (or quasi-2D) simulations for two main reasons. The first one was to avoid having mesh- to particle-size issues in the transverse direction, which would be problematic for computing accurately the fluid flow. The second reason is to allow measurements of the dynamics of the barchan central slice, something difficult to be done experimentally.

\section{Results and Discussion}

\subsection{Development of a dune from an initial heap}

\begin{figure}[h!]
	\begin{center}
		\includegraphics[width=.8\linewidth]{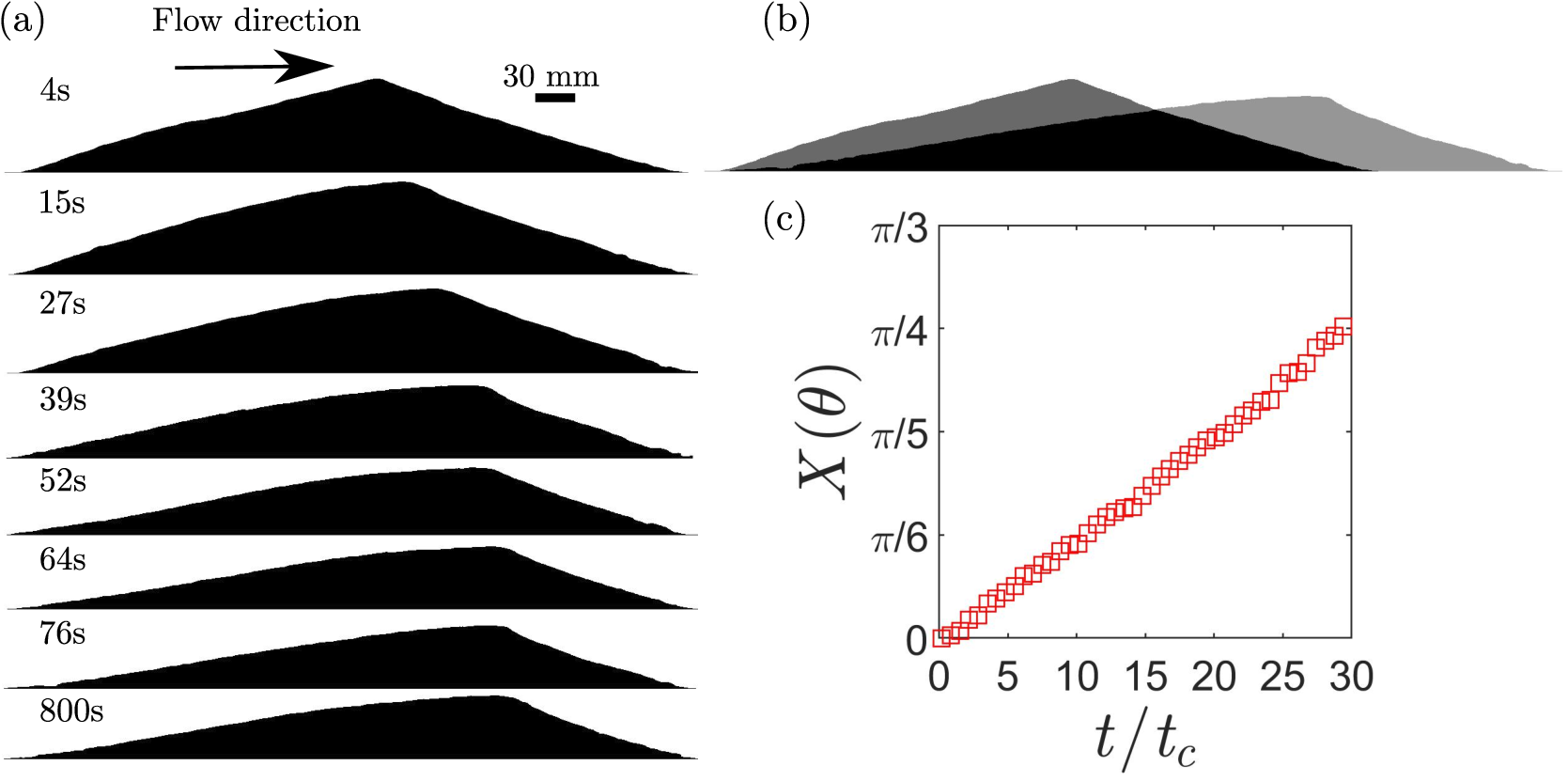}
	\end{center}
	\caption{\wa{(a) Snapshots showing lateral-view images of an initial heap being deformed into a 2D dune for case $c$ (Table \ref{flume_conditions}). The flow is from left to right in the images, and the corresponding time instants are shown on the left. (b) Superposition of the side view of the initial ($t$ = 4 s, in darker gray) and developed ($t$ = 76 s, in lighter gray) bedforms (intersection appears in black). The reference for the superposition was the crest position, shown in panel c. (c) Dune's displacement (based on the radian position \ef{in} the flume). The data are normalized by the timescale $t_c$\ef{, and $L$ $\approx$ 500 mm (corresponding to $\pi/6$) during the reversal process}.}}
	\label{fig:snaphots_2D}
\end{figure}

For the experiments outlined in cases \textit{a} to \textit{f} (Table \ref{flume_conditions}), we followed the bedform as it evolves from an initial heap into a 2D dune. For example, \ef{Figure \ref{fig:snaphots_2D}a} shows reconstructed snapshots of an initial pile being deformed into a 2D dune for case $c$. We initially observe the elongation of the upstream side and the formation of an avalanche face downstream of the crest, with the corresponding decrease of the crest height. Afterward, from a certain time on (76 s in this case), the dune keeps roughly the same shape, indicating a developed state. Figure \ref{fig:snaphots_2D}b shows the superposition of the side view of the initial ($t$ = 4 s, in darker gray) and developed ($t$ = 76 s, in lighter gray) dunes. If we consider that the intersected area (in black) is a good estimate of the number of grains that did not move (not necessarily equal, though), it indicates that a considerable part of the dune remains static (based on the superposed areas, 62\% of the dune remained static, with respect to the initially triangular pile), and that the dune reaches its developed form prior to a complete turnover. In order to investigate further the dune development, we measured the main morphological scales (length $L$, height $Z$ and slope $\theta$) along time.

\begin{figure}[h!]
	\begin{center}
		\includegraphics[width=1\linewidth]{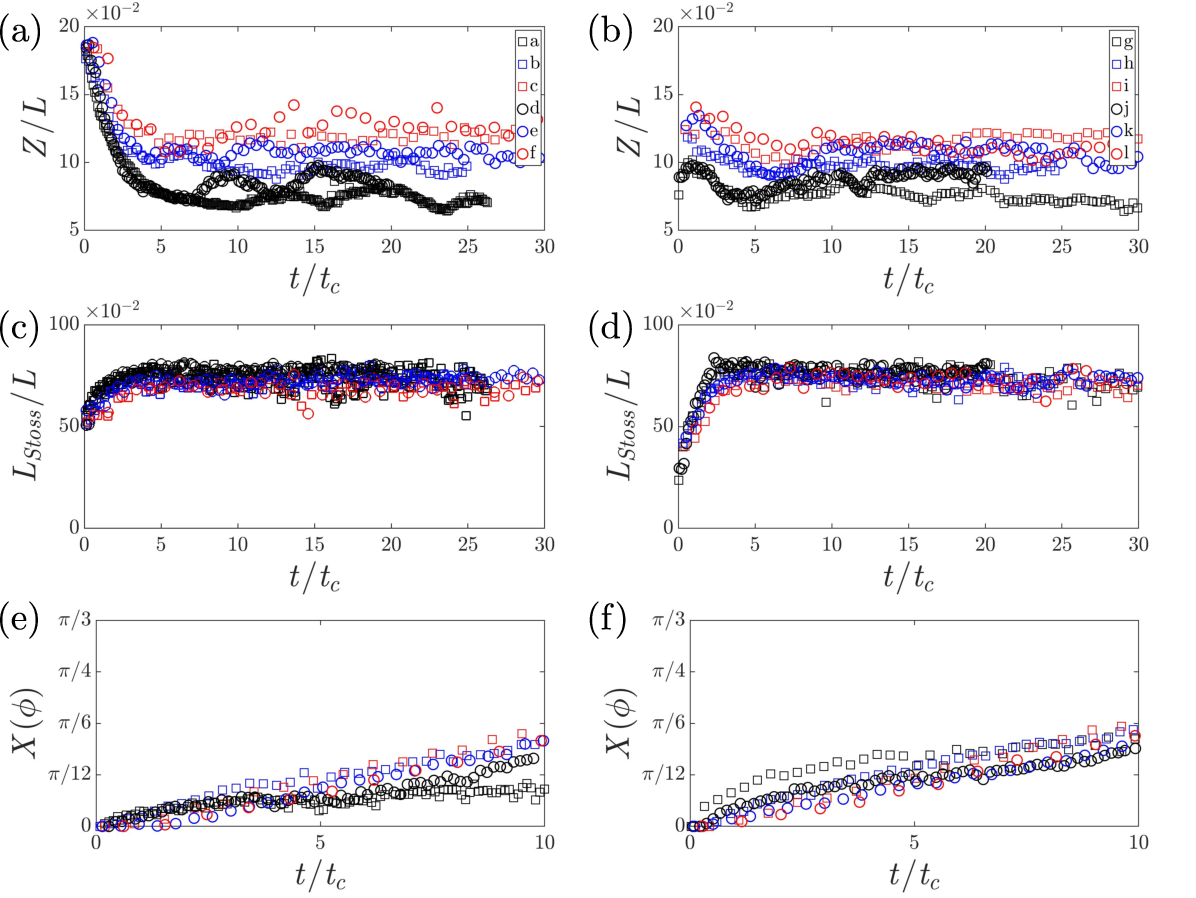}
	\end{center}
	\caption{\ef{(a) and (b) Time evolution of the vertical position of the maximum height (crest) of bedforms $Z$ normalized by the dune length $L$, for the initial development and reversal conditions, respectively. (c) and (d) Time evolution of the ratio of the length of the stoss side $L_{Stoss}$ to that of the entire dune $L$, for the initial development and reversal conditions, respectively. (e) and (f) Dune displacement during the initial development and flow reversal, respectively. In all panels, the time $t$ is normalized by the timescale $t_c$, and the figure keys in panels (a) and (b) refer to the tested conditions listed in Table \ref{flume_conditions}}.}
	\label{fig:Z_vs_t}
\end{figure}

Figure \ref{fig:Z_vs_t}a presents the vertical position of the maximum height (crest) of bedforms, $Z$, as a function of time, $t$ for the dune development\ef{. In this figure}, $Z$ is normalized by the total dune length in the streamwise direction, $L$, \ef{and the time by $t_{c}$,} which is a timescale for the growth of subaqueous barchans proposed by \citeA{Alvarez},

\begin{equation}
	t_{c} = \frac{L_{eq}(\rho_p/\rho_f)(\rho_p/\rho_f - 1)gd}{(u_*^2 - u_{th}^2)^{3/2}} \sim \frac{L_{eq}}{C} \,\,,
	\label{Eq:timescale}
\end{equation}

\noindent where $u_{th}$ is shear velocity at the threshold for the incipient motion of grains, $L_{eq}$ is the length of the developed dune, $g$ = $|\vec{g}|$, and

\begin{equation}
	C = \frac{q}{Z} \sim \frac{q}{L_{drag}} \,\,,
	\label{Eq:C}
\end{equation}

\noindent where $q$ is the transport rate, for which we considered the Meyer-Peter and M\"uller correlation \cite{Mueller}, and \wAssis{we assume that} the dune height $Z$ $\approx$ 0.1L \cite{Andreotti_1} is proportional to the saturation length $L_{sat}$, and, thus, to the drag length $L_{drag}$ = $(\rho_p/\rho_f)d$ \cite{Hersen_1}. We note that the inference $L$ $\sim$ $L_{sat}$ comes from stability analyses, which have shown that the most unstable mode of 
\wAssis{appearing dunes (that is, in their initial phase of development) scales} with the saturation length \cite{Andreotti_2, Claudin_Andreotti, Franklin_12, Franklin_5}, and in Equation \ref{Eq:timescale} we dropped all \wAssis{constants. In the nonlinear phase, the height of dunes can, in some cases, saturate while keeping the wavelength of the linear phase \cite{Franklin_5}, but in other cases the dune length can vary due to nonlinear processes, such as dune-dune interactions \cite{Assis}.} Because $t_c$ in Equation \ref{Eq:timescale} is proportional to $L_{eq}$ divided by the dune celerity C (displacement velocity of the dune crest), it scales with the dune turnover time ($t_c$ is a turnover time in which a transport-rate law was inserted, see the Supporting Information for more details on the expression for $t_c$). $t_c$ varied between 13 and 180 s in the experiments and 51 and 68 s in the simulations, values for each tested case are available in Tables S1 and S2 of supporting information. In Equation \ref{Eq:timescale}, we considered that $u_{th}$ = 0.007 m/s, \cite<in accordance with>{Andreotti_1}. In addition, the computed values have negligible effect in the ranges of $t_c$.

\begin{figure}[h!]
	\begin{center}
		\includegraphics[width=.90\linewidth]{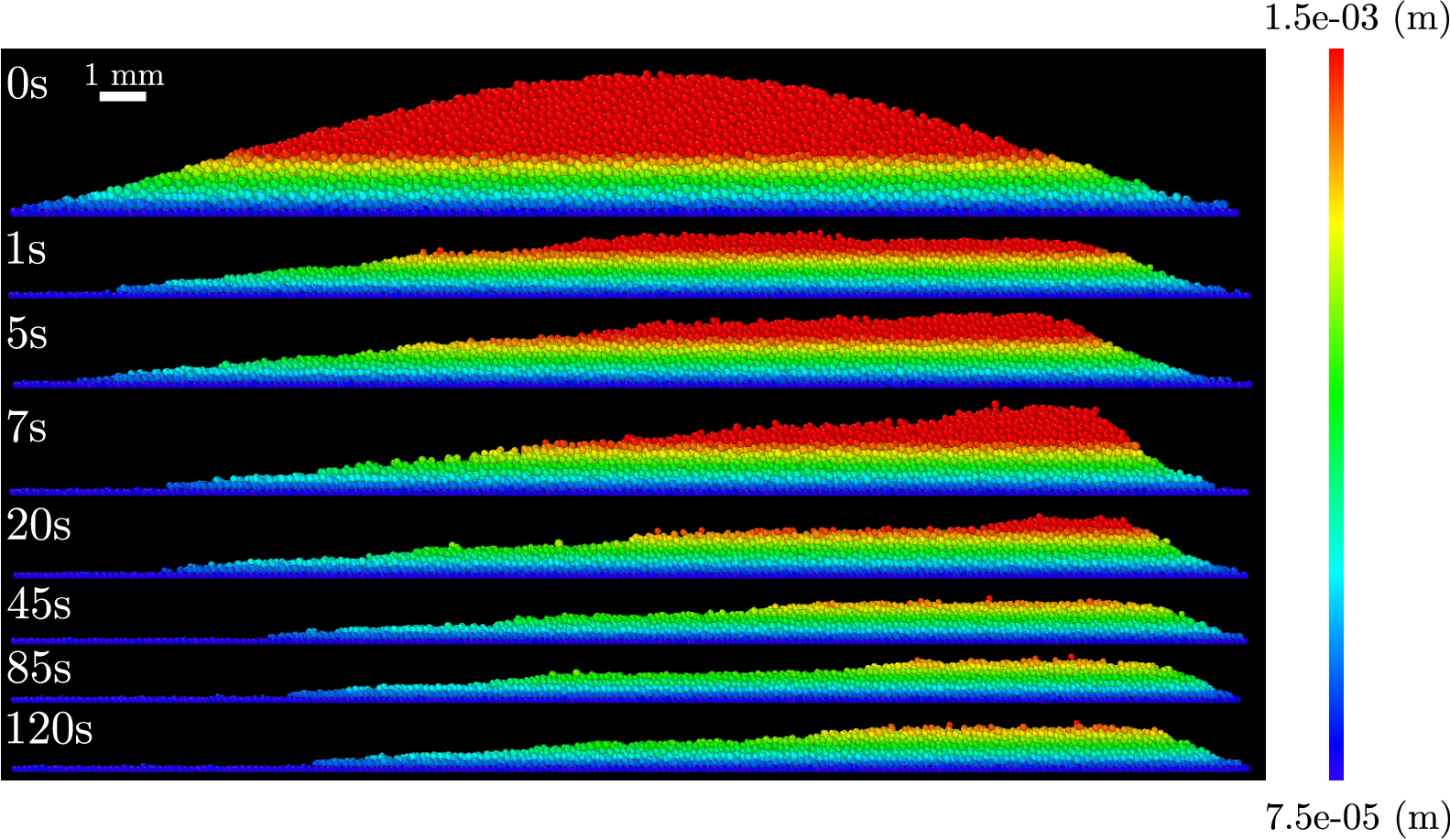}
	\end{center}
	\caption{Snapshots showing the central slice of a bedform being deformed into a barchan dune. The water flow is from left to right and the color represents the height (scale in the colorbar on the right). The corresponding time instants are shown on the left.}
	\label{fig:central_slice_one_direction}
\end{figure}

We observe that the aspect ratios showed in Figure 3(a) are different for different flow velocities, which confirms the theoretical models and experiments presented in previous studies \cite{Kroy_A, Parteli, Groh1}. For all cases, we observe in Figure \ref{fig:Z_vs_t} the existence of two characteristic times: a fast region occurring for $t/t_c$ $<$ 5, where $Z/L$ decreases relatively fast, and a slow one for $t/t_c$ $>$ 10, where $Z/L$ remains constant or oscillates around a mean value (plateau). While the fast time represents the flattening of the initial heap being deformed into a dune, the slow time indicates the presence of a developed dune. Therefore, the intersection between those two characteristic times corresponds to the typical time for the formation of a 2D dune from an initial heap, for which we find $t/t_c$ $\approx$ 5. Another way of searching for a characteristic time is through the evolution of the ratio of the stoss to total lengths, $L_{Stoss}/L$, which represents a relaxation toward the equilibrium shape (where $L_{Stoss}$ is the length of the stoss side, from the dune toe to its crest). For the initial development from a triangular pile, Figure \ref{fig:Z_vs_t}c shows that a plateau is reached at $t/t_c$ $\approx$ 5 for all tested cases, corroborating then the typical time proposed (see Figures S14 and S15 in the supporting information for fitted curves). Finally, Figure \ref{fig:Z_vs_t}e shows the dune displacement $X$, measured in terms of the channel angles. The value $\approx$ 5 is higher, but of the same order of magnitude, of that found by \citeA{Alvarez} for the development of barchan dunes based on the growth of their horns: $t/t_c$ $\approx$ 2.5. Because the mechanisms of barchan formation are different from those of 2D dunes, which do not have horns, this proximity of typical times is a strong indication of the existence of a similitude between the 2D dunes and the central slice of barchans. In order to inquire further into it, we performed three-dimensional CFD-DEM simulations of an initial pile being deformed into a barchan dune by a water flow, and analyze next the behavior of its central slice. Figure \ref{fig:central_slice_one_direction} shows snapshots of the central slice of a barchan dune (width equal to 2 mm, i.e., 10$d$) for different instants (see the supporting information for snapshots showing top view images of the barchan dune, and a movie showing the time evolution of the central slice). This width was chosen to avoid excessive fluctuations (due to the lack of grains in the spanwise direction) while analyzing the central slice only. Figures showing the longitudinal distribution of the slope, $\theta(x)$, for different time instants are available in the supporting information, for both the experiments and numerical simulations (central slice). They present a similar trend, with slightly higher mean values of $\theta(x)$ for the experiment.

\begin{figure}[h!]
	\begin{center}
		\includegraphics[width=.95\linewidth]{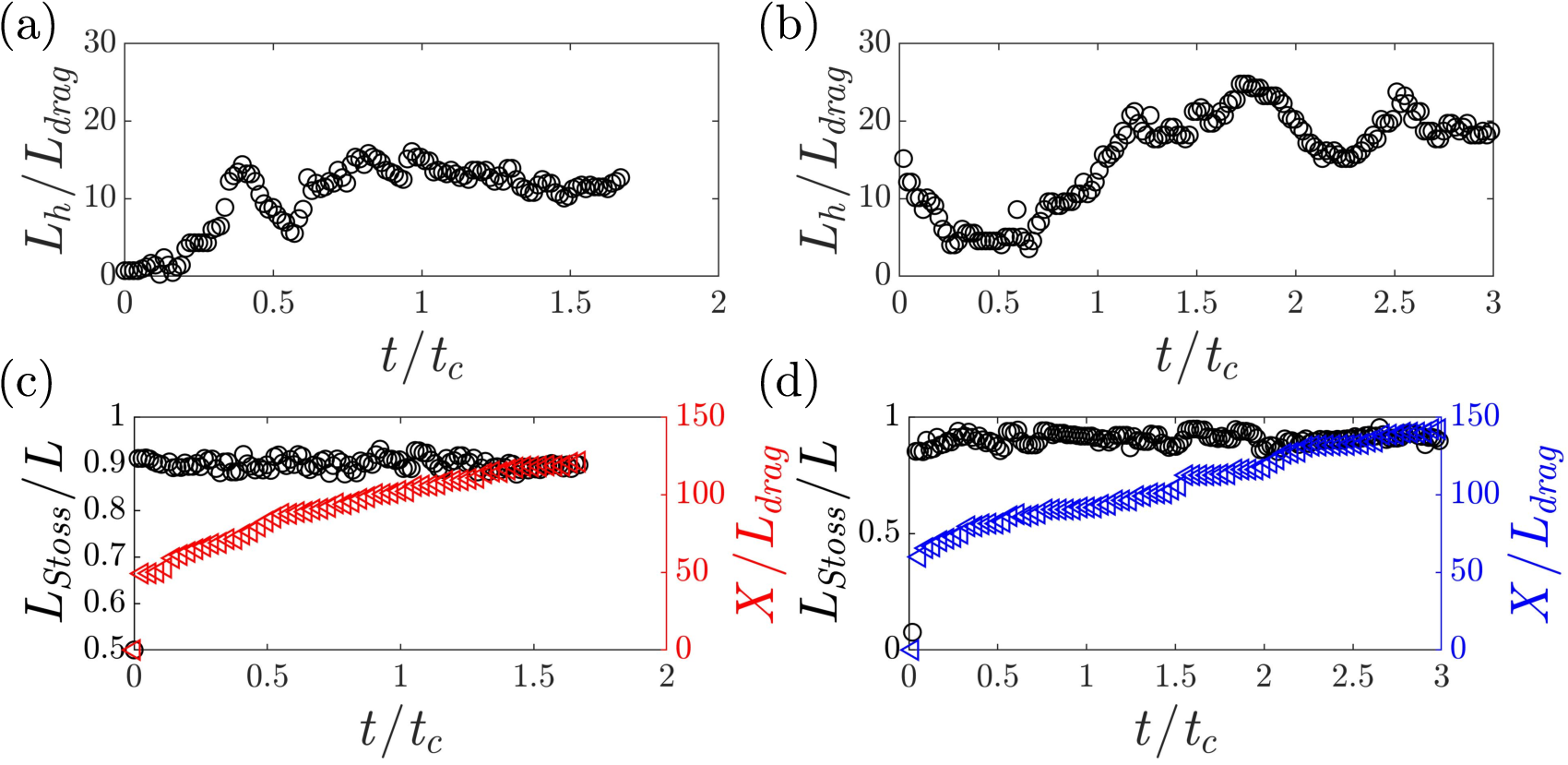}
	\end{center}
	\caption{(a) and (b) Evolution of the horn length $L_{horn}$ normalized by the characteristic length $L_{drag}$ for a barchan developed from a conical pile, and for a barchan undergoing flow reversal, respectively. (c) and (d) Time evolution of the ratio of the length of the stoss side $L_{Stoss}$ to that of the central slice $L$, and the dune displacement ($X$) normalized by the characteristic length $L_{drag}$ during the initial development and flow reversal, respectively. In all panels, the time $t$ is normalized by the timescale $t_c$.}
	\label{fig:data_horns}
\end{figure}

In our simulations, the central slice had a much smaller number of grains than the 2D dunes, which was imposed by the computational costs of the CFD-DEM simulations (we limited the total number of grains in order to keep simulation times small). Even with this size difference, we observe that Figure \ref{fig:central_slice_one_direction} shows a behavior similar to that of Figure \ref{fig:snaphots_2D}, with an elongation of the upstream side and formation of an avalanche face on the lee side, until a stable shape is reached (after 85 s. See figure S10 in the supporting information for the superposition of the central slice of the numerical dune). We note that the 3D bedform spreads laterally as the conical pile is deformed into a barchan dune, and Figure \ref{fig:central_slice_one_direction} shows the central slice only, giving the wrong impression that a large amount of grains was lost (see the supporting information for movies showing the time evolution of the central slice and of top view images of the entire dune). Figure \ref{fig:data_horns}a shows the time evolution of the horn length $L_h$, normalized by the timescale $L_{drag}$, as the conical pile is deformed into a barchan dune. In Figures \ref{fig:data_horns}a and \ref{fig:data_horns}b, $L_h$ is computed as the average of the two horns, and $L_{drag}$ is an inertial length, proportional to the flux saturation length \cite{Hersen_1, Andreotti_1, Andreotti_2, Claudin_Andreotti}. We observe an increase in $L_h$ along time, until a plateau is reached at $t/t_c$ $\approx$ 1--1.5, with $L_h$ oscillating around a mean value. The origin of oscillations are probably the small number of particles and the intermittent motion of grains. For this very small barchan, the time to reach the plateau is of the same order as that obtained experimentally by \citeA{Alvarez}. See in the supporting information (Figure S12) the data rescaled by the dune width $W$ and normalized by the turnover time $t_{turnover}$, the latter computed using the celerity measured directly from the dune celerity. It shows that $t_c$ is directly proportional to $t_{turnover}$, being a turnover time itself but without the necessity of knowing a priori the dune celerity.

Finally, Figures \ref{fig:data_horns}c and \ref{fig:data_horns}d show, respectively, the ratio of the length of the stoss side $L_{Stoss}$ to that of the entire dune $L$ and the dune displacement during the initial development. The time evolution of $L_{Stoss}/L$ for the 3D simulations is much more subtle than in the 2D experiments, and no conclusive remarks can be drawn from it. The same occurs for that of $Z/L$, which we show in Figure S11 of supporting information. We understand that comparisons between the temporal evolutions of barchan and 2D dunes must be in terms of horn size and $Z/L$, respectively, since grains experience significant transverse motions in the case of subaqueous barchans (leading to wrong interpretations if comparisons based on $Z/L$ only are adopted).

In summary, by comparing the formation of 2D dunes with that of barchans from an initial heap (triangular in two and conical in three dimensions), we observe a certain similarity between them, the central slice of the barchan dune behaving roughly as a 2D dune.

\subsection{Flow reversal}

\begin{figure}[h!]
	\begin{center}
		\includegraphics[width=.9\linewidth]{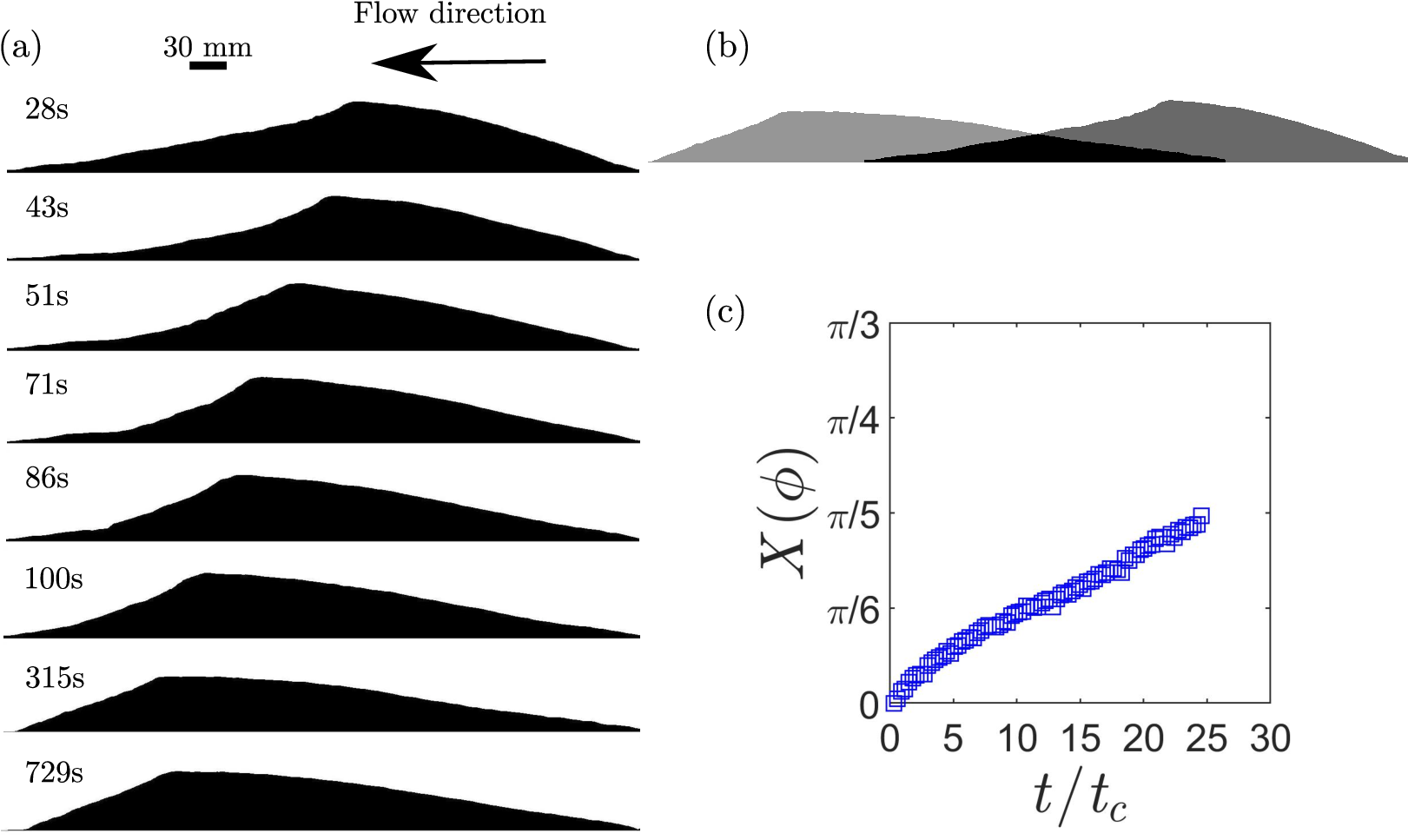}
	\end{center}
	\caption{(a) Snapshots showing lateral-view images of an initially developed 2D dune undergoing a flow reversal for case $h$ (Table \ref{flume_conditions}). The flow is from right to left in the images, and the corresponding time instants are shown on the left (time set to 0 s at the beginnig of the reversed flow). (b) Superposition of the side view of the initial ($t$ = 28 s, in darker gray) and developed ($t$ = 315 s, in lighter gray) bedforms (intersection appears in black). The reference for the superposition was the crest position\efb{, shown in panel c.} (c) Dune's displacement (based on the radian position in the flume). The data are normalized by the timescale $t_c$\, and $L$ $\approx$ 500 mm (corresponding to $\pi/6$) during the reversal process.}
	\label{fig:snapshots_2D_reversal}
\end{figure}

We inquire now into the process of inverting a dune by reversing the water flow. To create this condition, we performed experiments and numerical simulations in which we reversed the water flow after assuring that the dune was in a steady-state developed state. For the experiments with 2D dunes, this corresponds to cases \textit{g} to \textit{l} of Table \ref{flume_conditions}. Figure \ref{fig:snapshots_2D_reversal}a shows reconstructed snapshots of an initially developed 2D dune undergoing a flow reversal (case $h$). We notice that initially the motion occurs over the previous avalanche face, which has its slope decreased over time while the crest is displaced to the left. At the same time, a new lee face develops over the previous stoss side, with the crest and a small avalanche face migrating over it. During this process (within 28 s and 100 s in Figure \ref{fig:snapshots_2D_reversal}a), the new lee face has a varying angle, going from the avalanche angle (near the crest) to a very low slope (close to the new trailing edge). When the avalanche face reaches the trailing edge, the dune is properly inverted. Figure \ref{fig:snapshots_2D_reversal}b shows the superposition of the side view of the initial ($t$ = 28 s, in darker gray) and developed ($t$ = 315 s, in lighter gray) dunes. As for the development case, Figure \ref{fig:snapshots_2D_reversal}b indicates that a part of the dune remains static (if we consider that the intersected area is a good estimate of the static region, then $\approx$ 30\% of the dune remained static, with respect to the dune at the beginning of the flow reversal). In order to investigate further the reversal process, we measured the main morphological scales, which we present next.

Figure \ref{fig:Z_vs_t}b shows the vertical position of the maximum height (crest) of bedforms, $Z/L$, as a function of time, $t/t_c$, for cases \textit{g} to \textit{l}, respectively. We observe basically the existence of four characteristic times: (i) a fast region taking place in $t/t_c$ $<$ 1, in which $Z$ increases rapidly over time; (ii) a fast region occurring for 1 $<$ $t/t_c$ $<$ 5, for which $Z$ decreases \ef{rapidly} over time, representing the initial flattening of the dune. During the flattening, the crest region diffuses and moves downstream, and the former avalanche face moves over the former stoss slope (between 28 s and 71s in Figure \ref{fig:snapshots_2D_reversal}); (iii) another fast region occurring within 5 $<$ $t/t_c$ $<$ 10, for which $Z$ increases over time and presents lower slopes than in the first two regions. This is due to the formation of a new avalanche face over the former stoss side while the crest continues its downstream motion; and (iv) a slow region for $t/t_c$ $>$ 10, where $Z/L$ remains constant or oscillates around a mean value, indicating a developed form. Therefore, the total time for achieving an inverted dune is $t/t_c$ $\approx$ 10, approximately twice that for development from an initial heap. Figure \ref{fig:Z_vs_t}d shows the time evolution of $L_{Stoss}/L$, where the regions just described are represented by $L_{Stoss}/L$ slopes different from zero (although the slope for region 5 $<$ $t/t_c$ $<$ 10 is much more subtle than that for $t/t_c$ $<$ 5). Therefore, as in the case of the developing dune, $L_{Stoss}/L$ shows a plateau for $t/t_c$ $>$ 10, corroborating then the typical time proposed. Finally, Figure \ref{fig:Z_vs_t}f shows the dune displacement $X$, measured in terms of the channel angles.

We note that field evidences of reversing dunes were reported at the border between the Tibetan Plateau and the Taklamakan Desert, where dunes migrate in the opposite direction of that of the predicted (resultant) transport of sand. \citeA{Gao} showed that the migration direction is due to the significant speed up that occurs on the dune crest once the flow is reversed, generating a nonlinear motion of grains on the dune and a migration direction that depends on different slopes directly hit by storm events, explaining the apparent reverse motion. Although the flux of grains at the moment of the flow inversion is of importance, as shown by \citeA{Gao}, we did not measure it because of technical limitations in our experimental setup (it does not allow us to track individual grains).

\begin{figure}[h!]
	\begin{center}
		\includegraphics[width=.90\linewidth]{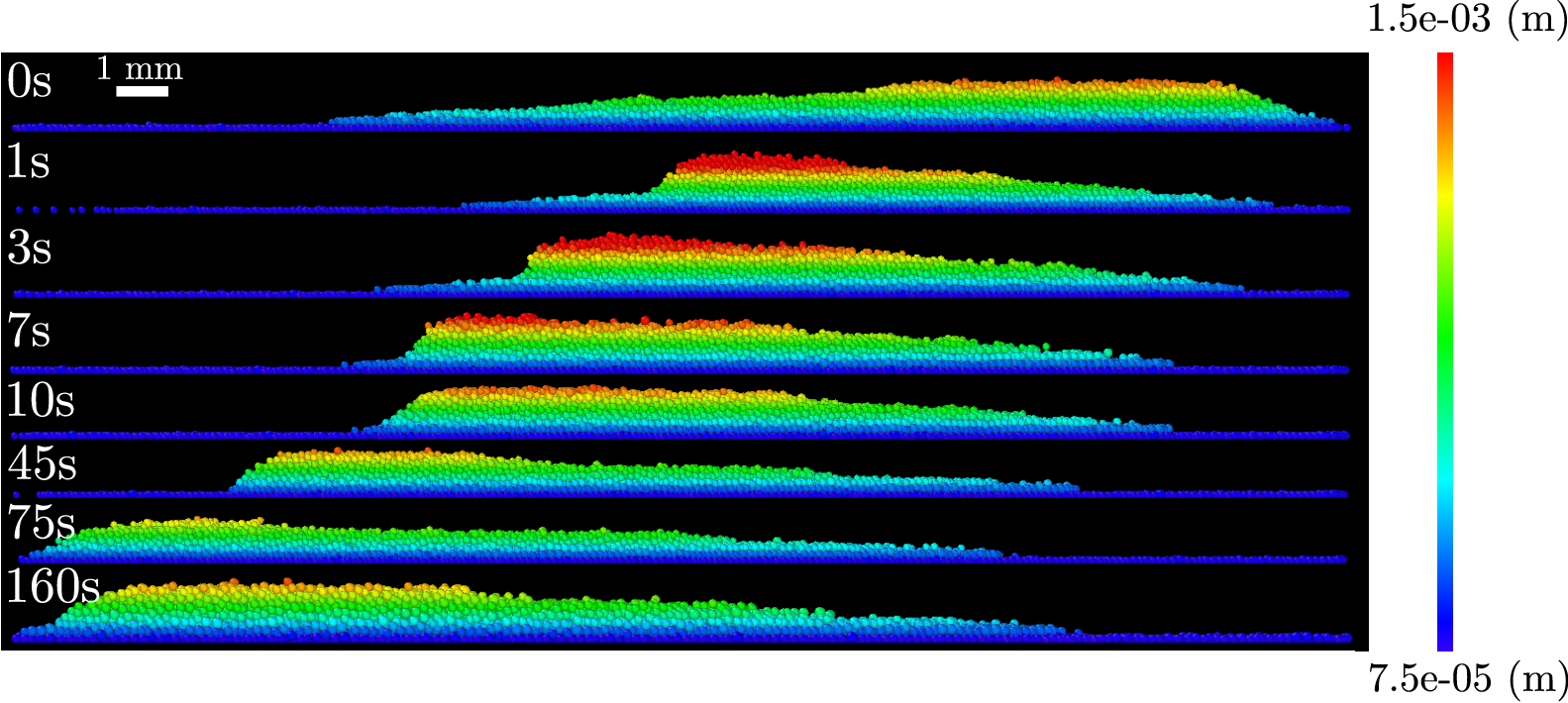}
	\end{center}
	\caption{Snapshots showing the central slice of a barchan dune undergoing a flow reversal. The water flow is from right to left, and the color represents the height (scale in the colorbar on the right). The corresponding time instants are shown on the left.}
	\label{fig:central_slice_reversal}
\end{figure}

Following a similar procedure as for the barchan formation, we carried out CFD-DEM simulations of a barchan undergoing inversion, and analyzed the behavior of its central slice. For that, we started with the developed barchan obtained in previous simulations and reversed the flow direction. Figure \ref{fig:central_slice_reversal} shows snapshots of the central slice of a barchan undergoing inversion for different instants (see the supporting information for snapshots showing top view images of the barchan dune, and a movie showing the central slice during the inversion). Although the central slice has a much smaller number of grains than the 2D dune, we observe a certain similarity between them: the crest and former avalanche face move over the former stoss side, and the latter becomes the new lee side. During the inversion process, the new lee side has a varying slope that goes from a very low angle (close to the new trailing edge, former toe) to an avalanche angle (just downstream the crest). The reversion process can be seen in more detail in Figure S13 of supporting information, which shows the snapshots of the central slice for the first 1 s after reversion has started. Figures showing $\theta(x)$ at different time instants for the reversing dune are available in the supporting information, for both the experiments and numerical simulations (central slice). Here, they also present a similar trend, with slightly higher mean values of $\theta(x)$ for the experiment.

In order to identify the time to attain a developed barchan, we proceeded as in \citeA{Alvarez} and tracked the growth of horns. Figure \ref{fig:data_horns}b shows the time evolution of the horn length $L_{h}$, normalized by the characteristic length $L_{drag}$, for a barchan undergoing reversal. We observe that initially the existing horns shrink ($L_{h}$ decreases), disappearing completely when $t/t_c$ $\approx$ 1, and from this time on the new horns begin to develop ($L_{h}$ increases). When  $t/t_c$ $\approx$ 2--2.5, the new horns seem to reach a developed state ($L_{h}$ reaches a plateau, oscillating around a mean value). Therefore, the barchan, as the 2D dune, takes twice the time to be completely inverted when compared with the formation from an initially conical heap. It is important to note that the characteristic times proposed can be different when barchans are formed from a flat bed, possibly leading to different formation times.

\subsection{Development vs reversal: $t_c$ and mobility of grains}

In the previous subsections, we compared 2D dunes with the central slice of 3D barchans. We found that the characteristic time for the development of 2D dunes is 5 $t_c$, where $t_c$ is a timescale used for the growth of barchan dunes.We also showed that for both 2D and barchan dunes the characteristic time to completely invert the dune under a flow reversal is twice that for the dune formation. These are relevant results indicating that the central slice of a barchan dune behaves roughly as a 2D dune, and they agree with previous studies. For instance, \citeA{Zhang_D} showed that the mean residence time of grains in the central slice of a barchan dune is around 10 turnover times (i.e., considerably large), the motion of grains being under the influence of lateral diffusion (outward motion) on the stoss slope and concentration (inward motion) in the center on the avalanche face. In this picture, together with the common timescale found in this work, the continuum models based on vertical slices that exchange mass between them are adequate for simulating barchan dunes. Those models are important for simulating large fields of barchans, for which the use of CFD-DEM is still prohibitive. In addition, because in Equation \ref{Eq:timescale} $t_c$ is computed using the transport rate proposed by \citeA{Mueller}, which has the same form of that proposed by \citeA{Bagnold_3} ($\sim$ $\theta^{3/2}$), we show that this kind of scaling law remains valid during reversals.

\begin{figure}[h!]
	\begin{center}
		\includegraphics[width=1\linewidth]{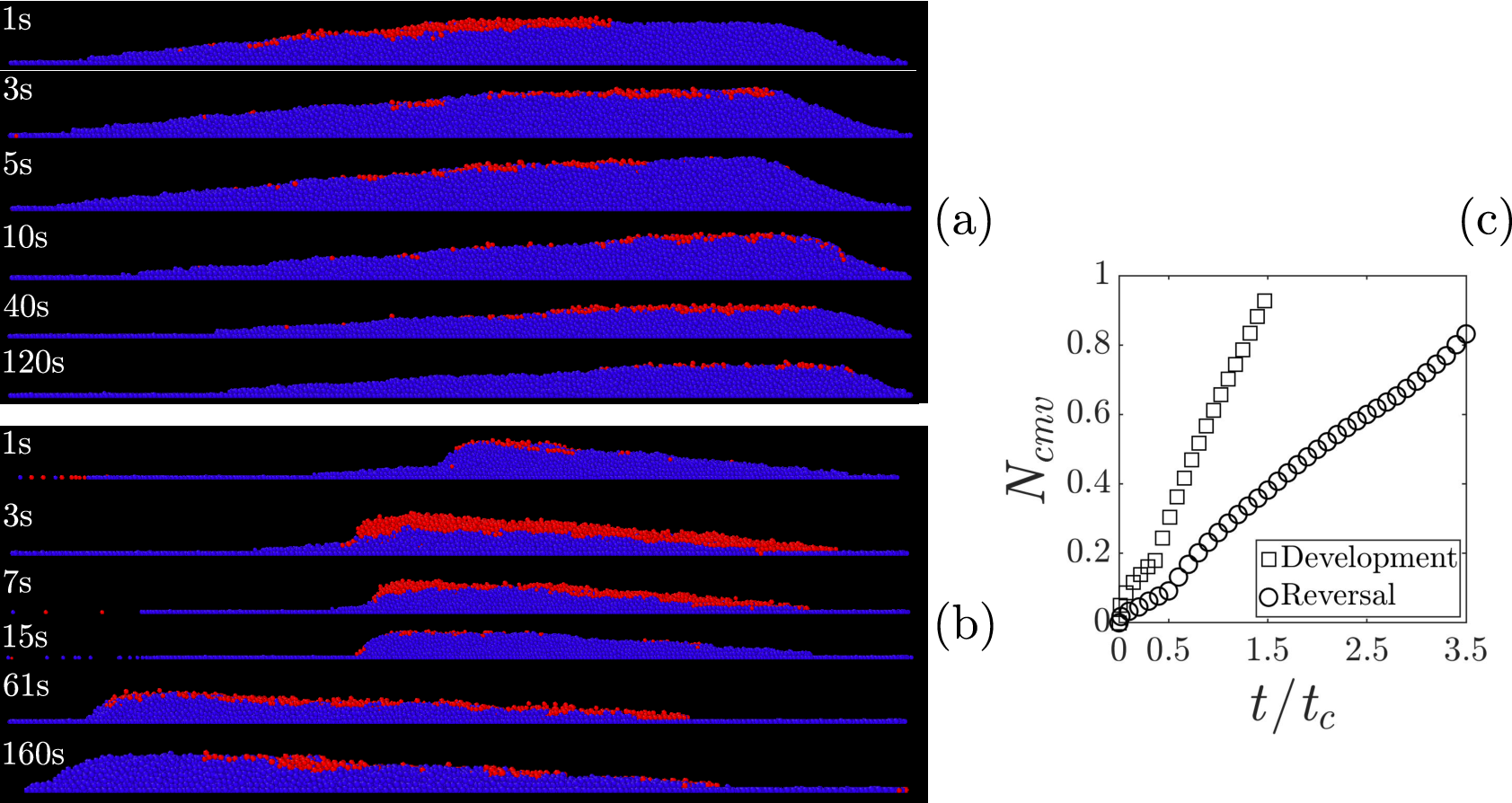}
	\end{center}
	\caption{Snapshots showing grains being entrained as bedload (red particles) and static (blue) in the central slice of a barchan dune. (a) Development from an initial heap and (b) barchan undergoing a reversal. The corresponding time instants are shown on the left. (c) Cumulative fraction of moving grains ($N_{cmv}$) moving in the central slice. The data are normalized by the timescale $t_c$.}
	\label{fig:central_slice_mobility}
\end{figure}

We now investigate the mobility of grains during the development and inversion of dunes. Since the numerical simulations compute the instantaneous position of each grain, we can track the motion of all grains as a function of time. Therefore, we measured the mobility of grains in the central slice during the barchan development and inversion. For example, Figure \ref{fig:central_slice_mobility} shows in red the grains with instantaneous velocities greater than 0.1$u_*$ \cite<typical bedload velocity over the dune,>{Wenzel}. We observe that only a few grains are mobilized within the central slice at each instant: only grains close to the surface move as bedload and grain below the surface remain static until exposed.

In order to know the proportion of moving grains with respect to the total, we counted the number of grains in the central slice that moved as bedload until a developed state was reached (we avoid counting the number of moving grains from 0 to 1 s due to the initialization of the flow). We obtained that approximately 23\% of grains remained static during the development from the initial heap and 20\% of grains remained static during the barchan reversal. We conclude that 1/5th of the grains in the central slice remain static when a dune develops from a different bedform. While the percentage of grains that remained static agrees roughly with that for the reversing 2D dunes, based on superposed areas, the percentage for the development from a conical (or triangular) heap does not agree. The latter disagreement can be due to inaccuracies when supposing that the superposition is proportional to the static region, and to the three dimensional motions in the barchan case as well. These differences remain, however, to be investigated further. A description of the procedure for identifying and counting the moving grains and a table listing the instantaneous number of grains moving as bedload at each instant are available in the supporting information.

We measured also the number of grains lost by the barchan dune along time. In terms of rates, we observed that during inversion the dune loses 10--15\% more grains than during its formation from a conical heap, as illustrated in the supporting information by tracing the amount of particles being lost over time.

Finally, we stress that the motion of grains is different in the eolian case (saltation, instead of rolling and sliding as in the subaqueous case), which results in much larger (hundreds of meters in length) and slower (years for the turnover time) scales. Therefore, any extrapolation of our results to the eolian case must be carried out with care.

\section{Conclusions}
In this paper, we investigated the similarities between a real 2D dune and the central slice of a 3D barchan dune, and how these dunes react under flow reversals. For that, we carried out experiments in a 2D flume and CFD-DEM simulations of 3D dunes where an initial heap was deformed into a dune that, by reversing the flow, evolved afterward toward an inverted dune. We found that the characteristic time for the development of 2D dunes is $\approx$ 5$t_c$, where $t_c$ is a timescale used for the growth of barchan dunes. By comparing with earlier work on 3D dunes, we concluded that the characteristic time-scale for 2D dunes is equivalent to that for 3D barchans. The reason is probably due to the fact that the central slice of barchans maintains a great part of its grains \cite<the outward motion in the transverse direction is balanced by inward avalanches, as shown by>{Zhang_D}. We showed that for both 2D and barchan dunes the characteristic time to completely invert the dune under a flow reversal is twice that for the dune formation, and we revealed the morphodynamics of reversing dunes: the grains on the lee side climb back the dune while its internal part and toe remain static, forming a new lee face. During the inversion process, the new lee side has a varying slope that goes from a very small angle (close to the new trailing edge, former toe) to an avalanche angle (just downstream the crest). We also showed that a considerable part of grains (around 20\%\, based on grain-scale simulations) remain static during the entire process, and that the barchan dune loses more grains during the reversal than during its formation from a conical pile. Our findings reveal the mechanisms for dune reversal, and provide a proof-of-concept that, in some cases, numerical similations of 3D barchans can be reduced to a central slice of a 2D dune, even in the subaqueous case.

%%%%%%%%%%%%%%%%%%%%%%%%%%%%%%%%%%%%%%%%%%%%%%%
\section*{Open Research}
\begin{sloppypar}
	Data (digital images) supporting this work were generated by ourselves and are available in Mendeley Data \cite{Supplemental2} under the CC-BY-4.0 license. The numerical scripts used to process the images and the numerical setup for simulations are also available in Mendeley Data \cite{Supplemental2} under the CC-BY-4.0 license.
\end{sloppypar}

\acknowledgments
Willian Assis and Erick Franklin are grateful to the S\~ao Paulo Research Foundation -- FAPESP (Grant Nos. 2018/14981-7, 2019/10239-7 and 2021/12910-8) and Conselho Nacional de Desenvolvimento Cient\'ifico e Tecnol\'ogico -- CNPq (Grant No. 405512/2022-8) for the financial support provided. N. M. Vriend is supported by a Royal Society University Research Fellowship No. URF/R1/191332. Willian Assis would like to express his gratitude to Karol Bacik and Nicolao Lima for their assistance in the experimental and numerical parts of this research, respectively.

\bibliography{references}

\begin{thebibliography}{}

\bibitem [\protect \citeauthoryear {%
Alvarez%
\ \BBA {} Franklin%
}{%
Alvarez%
\ \BBA {} Franklin%
}{%
{\protect \APACyear {2017}}%
}]{%
Alvarez}
\APACinsertmetastar {%
Alvarez}%
\begin{APACrefauthors}%
Alvarez, C\BPBI A.%
\BCBT {}\ \BBA {} Franklin, E\BPBI M.%
\end{APACrefauthors}%
\unskip\
\newblock
\APACrefYearMonthDay{2017}{Dec}{}.
\newblock
{\BBOQ}\APACrefatitle {Birth of a subaqueous barchan dune} {Birth of a
  subaqueous barchan dune}.{\BBCQ}
\newblock
\APACjournalVolNumPages{Phys. Rev. E}{96}{}{062906}.
\newblock
\begin{APACrefURL} \url{https://link.aps.org/doi/10.1103/PhysRevE.96.062906}
  \end{APACrefURL}
\newblock
\begin{APACrefDOI} \doi{10.1103/PhysRevE.96.062906} \end{APACrefDOI}
\PrintBackRefs{\CurrentBib}

\bibitem [\protect \citeauthoryear {%
Alvarez%
\ \BBA {} Franklin%
}{%
Alvarez%
\ \BBA {} Franklin%
}{%
{\protect \APACyear {2018}}%
}]{%
Alvarez3}
\APACinsertmetastar {%
Alvarez3}%
\begin{APACrefauthors}%
Alvarez, C\BPBI A.%
\BCBT {}\ \BBA {} Franklin, E\BPBI M.%
\end{APACrefauthors}%
\unskip\
\newblock
\APACrefYearMonthDay{2018}{Oct}{}.
\newblock
{\BBOQ}\APACrefatitle {Role of Transverse Displacements in the Formation of
  Subaqueous Barchan Dunes} {Role of transverse displacements in the formation
  of subaqueous barchan dunes}.{\BBCQ}
\newblock
\APACjournalVolNumPages{Phys. Rev. Lett.}{121}{}{164503}.
\newblock
\begin{APACrefURL}
  \url{https://link.aps.org/doi/10.1103/PhysRevLett.121.164503}
  \end{APACrefURL}
\newblock
\begin{APACrefDOI} \doi{10.1103/PhysRevLett.121.164503} \end{APACrefDOI}
\PrintBackRefs{\CurrentBib}

\bibitem [\protect \citeauthoryear {%
Alvarez%
\ \BBA {} Franklin%
}{%
Alvarez%
\ \BBA {} Franklin%
}{%
{\protect \APACyear {2019}}%
}]{%
Alvarez4}
\APACinsertmetastar {%
Alvarez4}%
\begin{APACrefauthors}%
Alvarez, C\BPBI A.%
\BCBT {}\ \BBA {} Franklin, E\BPBI M.%
\end{APACrefauthors}%
\unskip\
\newblock
\APACrefYearMonthDay{2019}{Oct}{}.
\newblock
{\BBOQ}\APACrefatitle {Horns of subaqueous barchan dunes: A study at the grain
  scale} {Horns of subaqueous barchan dunes: A study at the grain
  scale}.{\BBCQ}
\newblock
\APACjournalVolNumPages{Phys. Rev. E}{100}{}{042904}.
\newblock
\begin{APACrefURL} \url{https://link.aps.org/doi/10.1103/PhysRevE.100.042904}
  \end{APACrefURL}
\newblock
\begin{APACrefDOI} \doi{10.1103/PhysRevE.100.042904} \end{APACrefDOI}
\PrintBackRefs{\CurrentBib}

\bibitem [\protect \citeauthoryear {%
Alvarez%
\ \BBA {} Franklin%
}{%
Alvarez%
\ \BBA {} Franklin%
}{%
{\protect \APACyear {2020}}%
}]{%
Alvarez6}
\APACinsertmetastar {%
Alvarez6}%
\begin{APACrefauthors}%
Alvarez, C\BPBI A.%
\BCBT {}\ \BBA {} Franklin, E\BPBI M.%
\end{APACrefauthors}%
\unskip\
\newblock
\APACrefYearMonthDay{2020}{Jan}{}.
\newblock
{\BBOQ}\APACrefatitle {Shape evolution of numerically obtained subaqueous
  barchan dunes} {Shape evolution of numerically obtained subaqueous barchan
  dunes}.{\BBCQ}
\newblock
\APACjournalVolNumPages{Phys. Rev. E}{101}{}{012905}.
\newblock
\begin{APACrefURL} \url{https://link.aps.org/doi/10.1103/PhysRevE.101.012905}
  \end{APACrefURL}
\newblock
\begin{APACrefDOI} \doi{10.1103/PhysRevE.101.012905} \end{APACrefDOI}
\PrintBackRefs{\CurrentBib}

\bibitem [\protect \citeauthoryear {%
Alvarez%
\ \BBA {} Franklin%
}{%
Alvarez%
\ \BBA {} Franklin%
}{%
{\protect \APACyear {2021}}%
}]{%
Alvarez7}
\APACinsertmetastar {%
Alvarez7}%
\begin{APACrefauthors}%
Alvarez, C\BPBI A.%
\BCBT {}\ \BBA {} Franklin, E\BPBI M.%
\end{APACrefauthors}%
\unskip\
\newblock
\APACrefYearMonthDay{2021}{}{}.
\newblock
{\BBOQ}\APACrefatitle {Force distribution within a barchan dune} {Force
  distribution within a barchan dune}.{\BBCQ}
\newblock
\APACjournalVolNumPages{Phys. Fluids}{33}{1}{013313}.
\PrintBackRefs{\CurrentBib}

\bibitem [\protect \citeauthoryear {%
Andreotti%
, Claudin%
\BCBL {}\ \BBA {} Douady%
}{%
Andreotti%
\ \protect \BOthers {.}}{%
{\protect \APACyear {2002}}%
{\protect \APACexlab {{\protect \BCnt {1}}}}}]{%
Andreotti_1}
\APACinsertmetastar {%
Andreotti_1}%
\begin{APACrefauthors}%
Andreotti, B.%
, Claudin, P.%
\BCBL {}\ \BBA {} Douady, S.%
\end{APACrefauthors}%
\unskip\
\newblock
\APACrefYearMonthDay{2002{\protect \BCnt {1}}}{}{}.
\newblock
{\BBOQ}\APACrefatitle {Selection of dune shapes and velocities. Part 1:
  {D}ynamics of sand, wind and barchans} {Selection of dune shapes and
  velocities. part 1: {D}ynamics of sand, wind and barchans}.{\BBCQ}
\newblock
\APACjournalVolNumPages{Eur. Phys. J. B}{28}{}{321-329}.
\PrintBackRefs{\CurrentBib}

\bibitem [\protect \citeauthoryear {%
Andreotti%
, Claudin%
\BCBL {}\ \BBA {} Douady%
}{%
Andreotti%
\ \protect \BOthers {.}}{%
{\protect \APACyear {2002}}%
{\protect \APACexlab {{\protect \BCnt {2}}}}}]{%
Andreotti_2}
\APACinsertmetastar {%
Andreotti_2}%
\begin{APACrefauthors}%
Andreotti, B.%
, Claudin, P.%
\BCBL {}\ \BBA {} Douady, S.%
\end{APACrefauthors}%
\unskip\
\newblock
\APACrefYearMonthDay{2002{\protect \BCnt {2}}}{}{}.
\newblock
{\BBOQ}\APACrefatitle {Selection of dune shapes and velocities. Part 2: {A}
  two-dimensional model} {Selection of dune shapes and velocities. part 2: {A}
  two-dimensional model}.{\BBCQ}
\newblock
\APACjournalVolNumPages{Eur. Phys. J. B}{28}{}{341-352}.
\PrintBackRefs{\CurrentBib}

\bibitem [\protect \citeauthoryear {%
Andreotti%
, Fourri\`ere%
, Ould-Kaddour%
, Murray%
\BCBL {}\ \BBA {} Claudin%
}{%
Andreotti%
\ \protect \BOthers {.}}{%
{\protect \APACyear {2009}}%
}]{%
Andreotti_4}
\APACinsertmetastar {%
Andreotti_4}%
\begin{APACrefauthors}%
Andreotti, B.%
, Fourri\`ere, A.%
, Ould-Kaddour, F.%
, Murray, B.%
\BCBL {}\ \BBA {} Claudin, P.%
\end{APACrefauthors}%
\unskip\
\newblock
\APACrefYearMonthDay{2009}{}{}.
\newblock
{\BBOQ}\APACrefatitle {Giant aeolian dune size determined by the average depth
  of the atmospheric boundary layer} {Giant aeolian dune size determined by the
  average depth of the atmospheric boundary layer}.{\BBCQ}
\newblock
\APACjournalVolNumPages{Nature}{457}{}{1120-1123}.
\PrintBackRefs{\CurrentBib}

\bibitem [\protect \citeauthoryear {%
Assis%
, Borges%
\BCBL {}\ \BBA {} Franklin%
}{%
Assis%
, Borges%
\BCBL {}\ \BBA {} Franklin%
}{%
{\protect \APACyear {2023}}%
}]{%
Assis3}
\APACinsertmetastar {%
Assis3}%
\begin{APACrefauthors}%
Assis, W\BPBI R.%
, Borges, D\BPBI S.%
\BCBL {}\ \BBA {} Franklin, E\BPBI M.%
\end{APACrefauthors}%
\unskip\
\newblock
\APACrefYearMonthDay{2023}{}{}.
\newblock
{\BBOQ}\APACrefatitle {Barchan Dunes Cruising Dune-Size Obstacles} {Barchan
  dunes cruising dune-size obstacles}.{\BBCQ}
\newblock
\APACjournalVolNumPages{Geophys. Res. Lett.}{50}{14}{e2023GL104125}.
\newblock
\begin{APACrefURL}
  \url{https://agupubs.onlinelibrary.wiley.com/doi/abs/10.1029/2023GL104125}
  \end{APACrefURL}
\newblock
\begin{APACrefDOI} \doi{https://doi.org/10.1029/2023GL104125} \end{APACrefDOI}
\PrintBackRefs{\CurrentBib}

\bibitem [\protect \citeauthoryear {%
Assis%
\ \BBA {} Franklin%
}{%
Assis%
\ \BBA {} Franklin%
}{%
{\protect \APACyear {2020}}%
}]{%
Assis}
\APACinsertmetastar {%
Assis}%
\begin{APACrefauthors}%
Assis, W\BPBI R.%
\BCBT {}\ \BBA {} Franklin, E\BPBI M.%
\end{APACrefauthors}%
\unskip\
\newblock
\APACrefYearMonthDay{2020}{}{}.
\newblock
{\BBOQ}\APACrefatitle {A Comprehensive Picture for Binary Interactions of
  Subaqueous Barchans} {A comprehensive picture for binary interactions of
  subaqueous barchans}.{\BBCQ}
\newblock
\APACjournalVolNumPages{Geophys. Res. Lett.}{47}{18}{e2020GL089464}.
\PrintBackRefs{\CurrentBib}

\bibitem [\protect \citeauthoryear {%
Assis%
\ \BBA {} Franklin%
}{%
Assis%
\ \BBA {} Franklin%
}{%
{\protect \APACyear {2021}}%
}]{%
Assis2}
\APACinsertmetastar {%
Assis2}%
\begin{APACrefauthors}%
Assis, W\BPBI R.%
\BCBT {}\ \BBA {} Franklin, E\BPBI M.%
\end{APACrefauthors}%
\unskip\
\newblock
\APACrefYearMonthDay{2021}{}{}.
\newblock
{\BBOQ}\APACrefatitle {Morphodynamics of Barchan-Barchan Interactions
  Investigated at the Grain Scale} {Morphodynamics of barchan-barchan
  interactions investigated at the grain scale}.{\BBCQ}
\newblock
\APACjournalVolNumPages{J. Geophys. Res.: Earth Surf.}{126}{8}{e2021JF006237}.
\PrintBackRefs{\CurrentBib}

\bibitem [\protect \citeauthoryear {%
Assis%
, Franklin%
\BCBL {}\ \BBA {} Vriend%
}{%
Assis%
, Franklin%
\BCBL {}\ \BBA {} Vriend%
}{%
{\protect \APACyear {2023}}%
}]{%
Supplemental2}
\APACinsertmetastar {%
Supplemental2}%
\begin{APACrefauthors}%
Assis, W\BPBI R.%
, Franklin, E\BPBI M.%
\BCBL {}\ \BBA {} Vriend, N\BPBI M.%
\end{APACrefauthors}%
\unskip\
\newblock
\APACrefYearMonthDay{2023}{}{}.
\newblock
{\BBOQ}\APACrefatitle {Dataset for ``Evolving dunes under flow reversals: from
  an initial heap toward inverted dune'' [{D}ataset][{S}oftware]} {Dataset for
  ``evolving dunes under flow reversals: from an initial heap toward inverted
  dune'' [{D}ataset][{S}oftware]}.{\BBCQ}
\newblock
\APACjournalVolNumPages{Mendeley Data,
  http://dx.doi.org/10.17632/fw3bcrxknf.1}{}{}{}.
\newblock
\begin{APACrefDOI} \doi{10.17632/fw3bcrxknf.1} \end{APACrefDOI}
\PrintBackRefs{\CurrentBib}

\bibitem [\protect \citeauthoryear {%
Bacik%
, Canizares%
, Caulfield%
, Williams%
\BCBL {}\ \BBA {} Vriend%
}{%
Bacik%
, Canizares%
\BCBL {}\ \protect \BOthers {.}}{%
{\protect \APACyear {2021}}%
}]{%
bacik2021dynamics}
\APACinsertmetastar {%
bacik2021dynamics}%
\begin{APACrefauthors}%
Bacik, K\BPBI A.%
, Canizares, P.%
, Caulfield, C\BHBI c\BPBI P.%
, Williams, M\BPBI J.%
\BCBL {}\ \BBA {} Vriend, N\BPBI M.%
\end{APACrefauthors}%
\unskip\
\newblock
\APACrefYearMonthDay{2021}{Oct}{}.
\newblock
{\BBOQ}\APACrefatitle {Dynamics of migrating sand dunes interacting with
  obstacles} {Dynamics of migrating sand dunes interacting with
  obstacles}.{\BBCQ}
\newblock
\APACjournalVolNumPages{Phys. Rev. Fluids}{6}{}{104308}.
\newblock
\begin{APACrefURL}
  \url{https://link.aps.org/doi/10.1103/PhysRevFluids.6.104308}
  \end{APACrefURL}
\newblock
\begin{APACrefDOI} \doi{10.1103/PhysRevFluids.6.104308} \end{APACrefDOI}
\PrintBackRefs{\CurrentBib}

\bibitem [\protect \citeauthoryear {%
Bacik%
, Caulfield%
\BCBL {}\ \BBA {} Vriend%
}{%
Bacik%
, Caulfield%
\BCBL {}\ \BBA {} Vriend%
}{%
{\protect \APACyear {2021}}%
}]{%
bacik2021stability}
\APACinsertmetastar {%
bacik2021stability}%
\begin{APACrefauthors}%
Bacik, K\BPBI A.%
, Caulfield, C\BHBI c\BPBI P.%
\BCBL {}\ \BBA {} Vriend, N\BPBI M.%
\end{APACrefauthors}%
\unskip\
\newblock
\APACrefYearMonthDay{2021}{Oct}{}.
\newblock
{\BBOQ}\APACrefatitle {Stability of the Interaction between Two Sand Dunes in
  an Idealized Laboratory Experiment} {Stability of the interaction between two
  sand dunes in an idealized laboratory experiment}.{\BBCQ}
\newblock
\APACjournalVolNumPages{Phys. Rev. Lett.}{127}{}{154501}.
\newblock
\begin{APACrefURL}
  \url{https://link.aps.org/doi/10.1103/PhysRevLett.127.154501}
  \end{APACrefURL}
\newblock
\begin{APACrefDOI} \doi{10.1103/PhysRevLett.127.154501} \end{APACrefDOI}
\PrintBackRefs{\CurrentBib}

\bibitem [\protect \citeauthoryear {%
Bacik%
, Lovett%
, Caulfield%
\BCBL {}\ \BBA {} Vriend%
}{%
Bacik%
\ \protect \BOthers {.}}{%
{\protect \APACyear {2020}}%
}]{%
bacik2020wake}
\APACinsertmetastar {%
bacik2020wake}%
\begin{APACrefauthors}%
Bacik, K\BPBI A.%
, Lovett, S.%
, Caulfield, C\BHBI c\BPBI P.%
\BCBL {}\ \BBA {} Vriend, N\BPBI M.%
\end{APACrefauthors}%
\unskip\
\newblock
\APACrefYearMonthDay{2020}{Feb}{}.
\newblock
{\BBOQ}\APACrefatitle {Wake Induced Long Range Repulsion of Aqueous Dunes}
  {Wake induced long range repulsion of aqueous dunes}.{\BBCQ}
\newblock
\APACjournalVolNumPages{Phys. Rev. Lett.}{124}{}{054501}.
\newblock
\begin{APACrefURL}
  \url{https://link.aps.org/doi/10.1103/PhysRevLett.124.054501}
  \end{APACrefURL}
\newblock
\begin{APACrefDOI} \doi{10.1103/PhysRevLett.124.054501} \end{APACrefDOI}
\PrintBackRefs{\CurrentBib}

\bibitem [\protect \citeauthoryear {%
Bagnold%
}{%
Bagnold%
}{%
{\protect \APACyear {1941}}%
}]{%
Bagnold_1}
\APACinsertmetastar {%
Bagnold_1}%
\begin{APACrefauthors}%
Bagnold, R\BPBI A.%
\end{APACrefauthors}%
\unskip\
\newblock
\APACrefYear{1941}.
\newblock
\APACrefbtitle {The Physics of Blown Sand and Desert Dunes} {The physics of
  blown sand and desert dunes}.
\newblock
\APACaddressPublisher{London}{Chapman and Hall}.
\PrintBackRefs{\CurrentBib}

\bibitem [\protect \citeauthoryear {%
Bagnold%
}{%
Bagnold%
}{%
{\protect \APACyear {1956}}%
}]{%
Bagnold_3}
\APACinsertmetastar {%
Bagnold_3}%
\begin{APACrefauthors}%
Bagnold, R\BPBI A.%
\end{APACrefauthors}%
\unskip\
\newblock
\APACrefYearMonthDay{1956}{}{}.
\newblock
{\BBOQ}\APACrefatitle {The flow of cohesionless grains in fluids} {The flow of
  cohesionless grains in fluids}.{\BBCQ}
\newblock
\APACjournalVolNumPages{Philos. Trans. R. Soc. Lond. Ser. A}{249}{}{235-297}.
\PrintBackRefs{\CurrentBib}

\bibitem [\protect \citeauthoryear {%
Berger%
, Kloss%
, Kohlmeyer%
\BCBL {}\ \BBA {} Pirker%
}{%
Berger%
\ \protect \BOthers {.}}{%
{\protect \APACyear {2015}}%
}]{%
Berger}
\APACinsertmetastar {%
Berger}%
\begin{APACrefauthors}%
Berger, R.%
, Kloss, C.%
, Kohlmeyer, A.%
\BCBL {}\ \BBA {} Pirker, S.%
\end{APACrefauthors}%
\unskip\
\newblock
\APACrefYearMonthDay{2015}{}{}.
\newblock
{\BBOQ}\APACrefatitle {Hybrid parallelization of the {LIGGGHTS} open-source
  {DEM} code} {Hybrid parallelization of the {LIGGGHTS} open-source {DEM}
  code}.{\BBCQ}
\newblock
\APACjournalVolNumPages{Powder Technology}{278}{}{234-247}.
\PrintBackRefs{\CurrentBib}

\bibitem [\protect \citeauthoryear {%
Charru%
}{%
Charru%
}{%
{\protect \APACyear {2006}}%
}]{%
Charru_3}
\APACinsertmetastar {%
Charru_3}%
\begin{APACrefauthors}%
Charru, F.%
\end{APACrefauthors}%
\unskip\
\newblock
\APACrefYearMonthDay{2006}{}{}.
\newblock
{\BBOQ}\APACrefatitle {Selection of the ripple length on a granular bed sheared
  by a liquid flow} {Selection of the ripple length on a granular bed sheared
  by a liquid flow}.{\BBCQ}
\newblock
\APACjournalVolNumPages{Phys. Fluids}{18}{121508}{}.
\PrintBackRefs{\CurrentBib}

\bibitem [\protect \citeauthoryear {%
Claudin%
\ \BBA {} Andreotti%
}{%
Claudin%
\ \BBA {} Andreotti%
}{%
{\protect \APACyear {2006}}%
}]{%
Claudin_Andreotti}
\APACinsertmetastar {%
Claudin_Andreotti}%
\begin{APACrefauthors}%
Claudin, P.%
\BCBT {}\ \BBA {} Andreotti, B.%
\end{APACrefauthors}%
\unskip\
\newblock
\APACrefYearMonthDay{2006}{}{}.
\newblock
{\BBOQ}\APACrefatitle {A scaling law for aeolian dunes on {M}ars, {V}enus,
  {E}arth, and for subaqueous ripples} {A scaling law for aeolian dunes on
  {M}ars, {V}enus, {E}arth, and for subaqueous ripples}.{\BBCQ}
\newblock
\APACjournalVolNumPages{Earth Plan. Sci. Lett.}{252}{}{20-44}.
\PrintBackRefs{\CurrentBib}

\bibitem [\protect \citeauthoryear {%
{Courrech du Pont}%
}{%
{Courrech du Pont}%
}{%
{\protect \APACyear {2015}}%
}]{%
Courrech}
\APACinsertmetastar {%
Courrech}%
\begin{APACrefauthors}%
{Courrech du Pont}, S.%
\end{APACrefauthors}%
\unskip\
\newblock
\APACrefYearMonthDay{2015}{}{}.
\newblock
{\BBOQ}\APACrefatitle {Dune morphodynamics} {Dune morphodynamics}.{\BBCQ}
\newblock
\APACjournalVolNumPages{C. R. Phys.}{16}{1}{118 - 138}.
\PrintBackRefs{\CurrentBib}

\bibitem [\protect \citeauthoryear {%
Dur\'an%
, Schw\"ammle%
\BCBL {}\ \BBA {} Herrmann%
}{%
Dur\'an%
\ \protect \BOthers {.}}{%
{\protect \APACyear {2005}}%
}]{%
Duran2}
\APACinsertmetastar {%
Duran2}%
\begin{APACrefauthors}%
Dur\'an, O.%
, Schw\"ammle, V.%
\BCBL {}\ \BBA {} Herrmann, H.%
\end{APACrefauthors}%
\unskip\
\newblock
\APACrefYearMonthDay{2005}{Aug}{}.
\newblock
{\BBOQ}\APACrefatitle {Breeding and solitary wave behavior of dunes} {Breeding
  and solitary wave behavior of dunes}.{\BBCQ}
\newblock
\APACjournalVolNumPages{Phys. Rev. E}{72}{}{021308}.
\newblock
\begin{APACrefURL} \url{https://link.aps.org/doi/10.1103/PhysRevE.72.021308}
  \end{APACrefURL}
\newblock
\begin{APACrefDOI} \doi{10.1103/PhysRevE.72.021308} \end{APACrefDOI}
\PrintBackRefs{\CurrentBib}

\bibitem [\protect \citeauthoryear {%
Elbelrhiti%
, Claudin%
\BCBL {}\ \BBA {} Andreotti%
}{%
Elbelrhiti%
\ \protect \BOthers {.}}{%
{\protect \APACyear {2005}}%
}]{%
Elbelrhiti}
\APACinsertmetastar {%
Elbelrhiti}%
\begin{APACrefauthors}%
Elbelrhiti, H.%
, Claudin, P.%
\BCBL {}\ \BBA {} Andreotti, B.%
\end{APACrefauthors}%
\unskip\
\newblock
\APACrefYearMonthDay{2005}{}{}.
\newblock
{\BBOQ}\APACrefatitle {Field evidence for surface-wave-induced instability of
  sand dunes} {Field evidence for surface-wave-induced instability of sand
  dunes}.{\BBCQ}
\newblock
\APACjournalVolNumPages{Nature}{437}{04058}{}.
\PrintBackRefs{\CurrentBib}

\bibitem [\protect \citeauthoryear {%
Franklin%
}{%
Franklin%
}{%
{\protect \APACyear {2011}}%
}]{%
Franklin_5}
\APACinsertmetastar {%
Franklin_5}%
\begin{APACrefauthors}%
Franklin, E\BPBI M.%
\end{APACrefauthors}%
\unskip\
\newblock
\APACrefYearMonthDay{2011}{}{}.
\newblock
{\BBOQ}\APACrefatitle {Nonlinear instabilities on a granular bed sheared by a
  turbulent liquid flow} {Nonlinear instabilities on a granular bed sheared by
  a turbulent liquid flow}.{\BBCQ}
\newblock
\APACjournalVolNumPages{J. Braz. Soc. Mech. Sci. Eng.}{33}{}{265-271}.
\PrintBackRefs{\CurrentBib}

\bibitem [\protect \citeauthoryear {%
Franklin%
}{%
Franklin%
}{%
{\protect \APACyear {2015}}%
}]{%
Franklin_12}
\APACinsertmetastar {%
Franklin_12}%
\begin{APACrefauthors}%
Franklin, E\BPBI M.%
\end{APACrefauthors}%
\unskip\
\newblock
\APACrefYearMonthDay{2015}{}{}.
\newblock
{\BBOQ}\APACrefatitle {Formation of sand ripples under a turbulent liquid flow}
  {Formation of sand ripples under a turbulent liquid flow}.{\BBCQ}
\newblock
\APACjournalVolNumPages{Appl. Math. Model.}{39}{23}{7390-7400}.
\PrintBackRefs{\CurrentBib}

\bibitem [\protect \citeauthoryear {%
Franklin%
\ \BBA {} Charru%
}{%
Franklin%
\ \BBA {} Charru%
}{%
{\protect \APACyear {2011}}%
}]{%
Franklin_8}
\APACinsertmetastar {%
Franklin_8}%
\begin{APACrefauthors}%
Franklin, E\BPBI M.%
\BCBT {}\ \BBA {} Charru, F.%
\end{APACrefauthors}%
\unskip\
\newblock
\APACrefYearMonthDay{2011}{}{}.
\newblock
{\BBOQ}\APACrefatitle {Subaqueous barchan dunes in turbulent shear flow. {P}art
  1. {D}une motion} {Subaqueous barchan dunes in turbulent shear flow. {P}art
  1. {D}une motion}.{\BBCQ}
\newblock
\APACjournalVolNumPages{J. Fluid Mech.}{675}{}{199-222}.
\PrintBackRefs{\CurrentBib}

\bibitem [\protect \citeauthoryear {%
Gao%
, Narteau%
\BCBL {}\ \BBA {} Gadal%
}{%
Gao%
\ \protect \BOthers {.}}{%
{\protect \APACyear {2021}}%
}]{%
Gao}
\APACinsertmetastar {%
Gao}%
\begin{APACrefauthors}%
Gao, X.%
, Narteau, C.%
\BCBL {}\ \BBA {} Gadal, C.%
\end{APACrefauthors}%
\unskip\
\newblock
\APACrefYearMonthDay{2021}{}{}.
\newblock
{\BBOQ}\APACrefatitle {Migration of Reversing Dunes Against the Sand Flow Path
  as a Singular Expression of the Speed-Up Effect} {Migration of reversing
  dunes against the sand flow path as a singular expression of the speed-up
  effect}.{\BBCQ}
\newblock
\APACjournalVolNumPages{J. Geophys. Res.: Earth Surf.}{126}{5}{e2020JF005913}.
\newblock
\begin{APACrefURL}
  \url{https://agupubs.onlinelibrary.wiley.com/doi/abs/10.1029/2020JF005913}
  \end{APACrefURL}
\newblock
\begin{APACrefDOI} \doi{https://doi.org/10.1029/2020JF005913} \end{APACrefDOI}
\PrintBackRefs{\CurrentBib}

\bibitem [\protect \citeauthoryear {%
Goniva%
, Kloss%
, Deen%
, Kuipers%
\BCBL {}\ \BBA {} Pirker%
}{%
Goniva%
\ \protect \BOthers {.}}{%
{\protect \APACyear {2012}}%
}]{%
Goniva}
\APACinsertmetastar {%
Goniva}%
\begin{APACrefauthors}%
Goniva, C.%
, Kloss, C.%
, Deen, N\BPBI G.%
, Kuipers, J\BPBI A\BPBI M.%
\BCBL {}\ \BBA {} Pirker, S.%
\end{APACrefauthors}%
\unskip\
\newblock
\APACrefYearMonthDay{2012}{}{}.
\newblock
{\BBOQ}\APACrefatitle {Influence of rolling friction on single spout fluidized
  bed simulation} {Influence of rolling friction on single spout fluidized bed
  simulation}.{\BBCQ}
\newblock
\APACjournalVolNumPages{Particuology}{10}{5}{582-591}.
\PrintBackRefs{\CurrentBib}

\bibitem [\protect \citeauthoryear {%
Groh%
, Wierschem%
, Aksel%
, Rehberg%
\BCBL {}\ \BBA {} Kruelle%
}{%
Groh%
\ \protect \BOthers {.}}{%
{\protect \APACyear {2008}}%
}]{%
Groh1}
\APACinsertmetastar {%
Groh1}%
\begin{APACrefauthors}%
Groh, C.%
, Wierschem, A.%
, Aksel, N.%
, Rehberg, I.%
\BCBL {}\ \BBA {} Kruelle, C\BPBI A.%
\end{APACrefauthors}%
\unskip\
\newblock
\APACrefYearMonthDay{2008}{Aug}{}.
\newblock
{\BBOQ}\APACrefatitle {Barchan dunes in two dimensions: Experimental tests for
  minimal models} {Barchan dunes in two dimensions: Experimental tests for
  minimal models}.{\BBCQ}
\newblock
\APACjournalVolNumPages{Phys. Rev. E}{78}{}{021304}.
\newblock
\begin{APACrefURL} \url{https://link.aps.org/doi/10.1103/PhysRevE.78.021304}
  \end{APACrefURL}
\newblock
\begin{APACrefDOI} \doi{10.1103/PhysRevE.78.021304} \end{APACrefDOI}
\PrintBackRefs{\CurrentBib}

\bibitem [\protect \citeauthoryear {%
Guignier%
, Niiya%
, Nishimori%
, Lague%
\BCBL {}\ \BBA {} Valance%
}{%
Guignier%
\ \protect \BOthers {.}}{%
{\protect \APACyear {2013}}%
}]{%
Guignier}
\APACinsertmetastar {%
Guignier}%
\begin{APACrefauthors}%
Guignier, L.%
, Niiya, H.%
, Nishimori, H.%
, Lague, D.%
\BCBL {}\ \BBA {} Valance, A.%
\end{APACrefauthors}%
\unskip\
\newblock
\APACrefYearMonthDay{2013}{May}{}.
\newblock
{\BBOQ}\APACrefatitle {Sand dunes as migrating strings} {Sand dunes as
  migrating strings}.{\BBCQ}
\newblock
\APACjournalVolNumPages{Phys. Rev. E}{87}{}{052206}.
\newblock
\begin{APACrefURL} \url{https://link.aps.org/doi/10.1103/PhysRevE.87.052206}
  \end{APACrefURL}
\newblock
\begin{APACrefDOI} \doi{10.1103/PhysRevE.87.052206} \end{APACrefDOI}
\PrintBackRefs{\CurrentBib}

\bibitem [\protect \citeauthoryear {%
Herrmann%
\ \BBA {} Sauermann%
}{%
Herrmann%
\ \BBA {} Sauermann%
}{%
{\protect \APACyear {2000}}%
}]{%
Herrmann_Sauermann}
\APACinsertmetastar {%
Herrmann_Sauermann}%
\begin{APACrefauthors}%
Herrmann, H\BPBI J.%
\BCBT {}\ \BBA {} Sauermann, G.%
\end{APACrefauthors}%
\unskip\
\newblock
\APACrefYearMonthDay{2000}{}{}.
\newblock
{\BBOQ}\APACrefatitle {The shape of dunes} {The shape of dunes}.{\BBCQ}
\newblock
\APACjournalVolNumPages{Physica A (Amsterdam)}{283}{}{24-30}.
\PrintBackRefs{\CurrentBib}

\bibitem [\protect \citeauthoryear {%
Hersen%
}{%
Hersen%
}{%
{\protect \APACyear {2004}}%
}]{%
Hersen_3}
\APACinsertmetastar {%
Hersen_3}%
\begin{APACrefauthors}%
Hersen, P.%
\end{APACrefauthors}%
\unskip\
\newblock
\APACrefYearMonthDay{2004}{}{}.
\newblock
{\BBOQ}\APACrefatitle {On the crescentic shape of barchan dunes} {On the
  crescentic shape of barchan dunes}.{\BBCQ}
\newblock
\APACjournalVolNumPages{Eur. Phys. J. B}{37}{4}{507--514}.
\PrintBackRefs{\CurrentBib}

\bibitem [\protect \citeauthoryear {%
Hersen%
\ \protect \BOthers {.}}{%
Hersen%
\ \protect \BOthers {.}}{%
{\protect \APACyear {2004}}%
}]{%
Hersen_2}
\APACinsertmetastar {%
Hersen_2}%
\begin{APACrefauthors}%
Hersen, P.%
, Andersen, K\BPBI H.%
, Elbelrhiti, H.%
, Andreotti, B.%
, Claudin, P.%
\BCBL {}\ \BBA {} Douady, S.%
\end{APACrefauthors}%
\unskip\
\newblock
\APACrefYearMonthDay{2004}{Jan}{}.
\newblock
{\BBOQ}\APACrefatitle {Corridors of barchan dunes: Stability and size
  selection} {Corridors of barchan dunes: Stability and size selection}.{\BBCQ}
\newblock
\APACjournalVolNumPages{Phys. Rev. E}{69}{}{011304}.
\newblock
\begin{APACrefURL} \url{https://link.aps.org/doi/10.1103/PhysRevE.69.011304}
  \end{APACrefURL}
\newblock
\begin{APACrefDOI} \doi{10.1103/PhysRevE.69.011304} \end{APACrefDOI}
\PrintBackRefs{\CurrentBib}

\bibitem [\protect \citeauthoryear {%
Hersen%
, Douady%
\BCBL {}\ \BBA {} Andreotti%
}{%
Hersen%
\ \protect \BOthers {.}}{%
{\protect \APACyear {2002}}%
}]{%
Hersen_1}
\APACinsertmetastar {%
Hersen_1}%
\begin{APACrefauthors}%
Hersen, P.%
, Douady, S.%
\BCBL {}\ \BBA {} Andreotti, B.%
\end{APACrefauthors}%
\unskip\
\newblock
\APACrefYearMonthDay{2002}{Dec}{}.
\newblock
{\BBOQ}\APACrefatitle {Relevant Length Scale of Barchan Dunes} {Relevant length
  scale of barchan dunes}.{\BBCQ}
\newblock
\APACjournalVolNumPages{Phys. Rev. Lett.}{89}{}{264301}.
\newblock
\begin{APACrefURL} \url{https://link.aps.org/doi/10.1103/PhysRevLett.89.264301}
  \end{APACrefURL}
\newblock
\begin{APACrefDOI} \doi{10.1103/PhysRevLett.89.264301} \end{APACrefDOI}
\PrintBackRefs{\CurrentBib}

\bibitem [\protect \citeauthoryear {%
Jarvis%
, Bacik%
, Narteau%
\BCBL {}\ \BBA {} Vriend%
}{%
Jarvis%
\ \protect \BOthers {.}}{%
{\protect \APACyear {2022}}%
}]{%
jarvis2022coarsening}
\APACinsertmetastar {%
jarvis2022coarsening}%
\begin{APACrefauthors}%
Jarvis, P.%
, Bacik, K.%
, Narteau, C.%
\BCBL {}\ \BBA {} Vriend, N.%
\end{APACrefauthors}%
\unskip\
\newblock
\APACrefYearMonthDay{2022}{}{}.
\newblock
{\BBOQ}\APACrefatitle {Coarsening dynamics of {2D} subaqueous dunes}
  {Coarsening dynamics of {2D} subaqueous dunes}.{\BBCQ}
\newblock
\APACjournalVolNumPages{Journal of Geophysical Research: Earth
  Surface}{127}{2}{e2021JF006492}.
\PrintBackRefs{\CurrentBib}

\bibitem [\protect \citeauthoryear {%
Khosronejad%
\ \BBA {} Sotiropoulos%
}{%
Khosronejad%
\ \BBA {} Sotiropoulos%
}{%
{\protect \APACyear {2017}}%
}]{%
Khosronejad}
\APACinsertmetastar {%
Khosronejad}%
\begin{APACrefauthors}%
Khosronejad, A.%
\BCBT {}\ \BBA {} Sotiropoulos, F.%
\end{APACrefauthors}%
\unskip\
\newblock
\APACrefYearMonthDay{2017}{}{}.
\newblock
{\BBOQ}\APACrefatitle {On the genesis and evolution of barchan dunes:
  morphodynamics} {On the genesis and evolution of barchan dunes:
  morphodynamics}.{\BBCQ}
\newblock
\APACjournalVolNumPages{J. Fluid Mech.}{815}{}{117--148}.
\PrintBackRefs{\CurrentBib}

\bibitem [\protect \citeauthoryear {%
Kidanemariam%
\ \BBA {} Uhlmann%
}{%
Kidanemariam%
\ \BBA {} Uhlmann%
}{%
{\protect \APACyear {2014}}%
}]{%
Kidanemariam}
\APACinsertmetastar {%
Kidanemariam}%
\begin{APACrefauthors}%
Kidanemariam, A\BPBI G.%
\BCBT {}\ \BBA {} Uhlmann, M.%
\end{APACrefauthors}%
\unskip\
\newblock
\APACrefYearMonthDay{2014}{}{}.
\newblock
{\BBOQ}\APACrefatitle {Direct numerical simulation of pattern formation in
  subaqueous sediment} {Direct numerical simulation of pattern formation in
  subaqueous sediment}.{\BBCQ}
\newblock
\APACjournalVolNumPages{J. Fluid Mech.}{750}{}{R2}.
\PrintBackRefs{\CurrentBib}

\bibitem [\protect \citeauthoryear {%
Kloss%
\ \BBA {} Goniva%
}{%
Kloss%
\ \BBA {} Goniva%
}{%
{\protect \APACyear {2010}}%
}]{%
Kloss}
\APACinsertmetastar {%
Kloss}%
\begin{APACrefauthors}%
Kloss, C.%
\BCBT {}\ \BBA {} Goniva, C.%
\end{APACrefauthors}%
\unskip\
\newblock
\APACrefYearMonthDay{2010}{}{}.
\newblock
{\BBOQ}\APACrefatitle {{LIGGGHTS}: a new open source discrete element
  simulation software} {{LIGGGHTS}: a new open source discrete element
  simulation software}.{\BBCQ}
\newblock
\BIn{} \APACrefbtitle {Proc. 5th Int. Conf. on Discrete Element Methods.}
  {Proc. 5th int. conf. on discrete element methods.}
\newblock
\APACaddressPublisher{London, UK}{}.
\PrintBackRefs{\CurrentBib}

\bibitem [\protect \citeauthoryear {%
Kroy%
, Fischer%
\BCBL {}\ \BBA {} Obermayer%
}{%
Kroy%
\ \protect \BOthers {.}}{%
{\protect \APACyear {2005}}%
}]{%
Kroy_B}
\APACinsertmetastar {%
Kroy_B}%
\begin{APACrefauthors}%
Kroy, K.%
, Fischer, S.%
\BCBL {}\ \BBA {} Obermayer, B.%
\end{APACrefauthors}%
\unskip\
\newblock
\APACrefYearMonthDay{2005}{}{}.
\newblock
{\BBOQ}\APACrefatitle {The shape of barchan dunes} {The shape of barchan
  dunes}.{\BBCQ}
\newblock
\APACjournalVolNumPages{J. Phys. Condens. Matter}{17}{}{S1229-0S1235}.
\PrintBackRefs{\CurrentBib}

\bibitem [\protect \citeauthoryear {%
Kroy%
, Sauermann%
\BCBL {}\ \BBA {} Herrmann%
}{%
Kroy%
\ \protect \BOthers {.}}{%
{\protect \APACyear {2002}}%
{\protect \APACexlab {{\protect \BCnt {1}}}}}]{%
Kroy_A}
\APACinsertmetastar {%
Kroy_A}%
\begin{APACrefauthors}%
Kroy, K.%
, Sauermann, G.%
\BCBL {}\ \BBA {} Herrmann, H\BPBI J.%
\end{APACrefauthors}%
\unskip\
\newblock
\APACrefYearMonthDay{2002{\protect \BCnt {1}}}{Sep}{}.
\newblock
{\BBOQ}\APACrefatitle {Minimal model for aeolian sand dunes} {Minimal model for
  aeolian sand dunes}.{\BBCQ}
\newblock
\APACjournalVolNumPages{Phys. Rev. E}{66}{}{031302}.
\newblock
\begin{APACrefURL} \url{https://link.aps.org/doi/10.1103/PhysRevE.66.031302}
  \end{APACrefURL}
\newblock
\begin{APACrefDOI} \doi{10.1103/PhysRevE.66.031302} \end{APACrefDOI}
\PrintBackRefs{\CurrentBib}

\bibitem [\protect \citeauthoryear {%
Kroy%
, Sauermann%
\BCBL {}\ \BBA {} Herrmann%
}{%
Kroy%
\ \protect \BOthers {.}}{%
{\protect \APACyear {2002}}%
{\protect \APACexlab {{\protect \BCnt {2}}}}}]{%
Kroy_C}
\APACinsertmetastar {%
Kroy_C}%
\begin{APACrefauthors}%
Kroy, K.%
, Sauermann, G.%
\BCBL {}\ \BBA {} Herrmann, H\BPBI J.%
\end{APACrefauthors}%
\unskip\
\newblock
\APACrefYearMonthDay{2002{\protect \BCnt {2}}}{Jan}{}.
\newblock
{\BBOQ}\APACrefatitle {Minimal Model for Sand Dunes} {Minimal model for sand
  dunes}.{\BBCQ}
\newblock
\APACjournalVolNumPages{Phys. Rev. Lett.}{88}{}{054301}.
\newblock
\begin{APACrefURL} \url{https://link.aps.org/doi/10.1103/PhysRevLett.88.054301}
  \end{APACrefURL}
\newblock
\begin{APACrefDOI} \doi{10.1103/PhysRevLett.88.054301} \end{APACrefDOI}
\PrintBackRefs{\CurrentBib}

\bibitem [\protect \citeauthoryear {%
Lima%
, Assis%
, Alvarez%
\BCBL {}\ \BBA {} Franklin%
}{%
Lima%
\ \protect \BOthers {.}}{%
{\protect \APACyear {2022}}%
}]{%
Lima2}
\APACinsertmetastar {%
Lima2}%
\begin{APACrefauthors}%
Lima, N\BPBI C.%
, Assis, W\BPBI R.%
, Alvarez, C\BPBI A.%
\BCBL {}\ \BBA {} Franklin, E\BPBI M.%
\end{APACrefauthors}%
\unskip\
\newblock
\APACrefYearMonthDay{2022}{}{}.
\newblock
{\BBOQ}\APACrefatitle {Grain-scale computations of barchan dunes} {Grain-scale
  computations of barchan dunes}.{\BBCQ}
\newblock
\APACjournalVolNumPages{Phys. Fluids}{34}{12}{123320}.
\newblock
\begin{APACrefURL} \url{https://doi.org/10.1063/5.0121810} \end{APACrefURL}
\newblock
\begin{APACrefDOI} \doi{10.1063/5.0121810} \end{APACrefDOI}
\PrintBackRefs{\CurrentBib}

\bibitem [\protect \citeauthoryear {%
Meyer-Peter%
\ \BBA {} M\"{u}ller%
}{%
Meyer-Peter%
\ \BBA {} M\"{u}ller%
}{%
{\protect \APACyear {1948}}%
}]{%
Mueller}
\APACinsertmetastar {%
Mueller}%
\begin{APACrefauthors}%
Meyer-Peter, E.%
\BCBT {}\ \BBA {} M\"{u}ller, R.%
\end{APACrefauthors}%
\unskip\
\newblock
\APACrefYearMonthDay{1948}{}{}.
\newblock
{\BBOQ}\APACrefatitle {Formulas for bed-load transport} {Formulas for bed-load
  transport}.{\BBCQ}
\newblock
\BIn{} \APACrefbtitle {Proc. 2nd Meeting of International Association for
  Hydraulic Research} {Proc. 2nd meeting of international association for
  hydraulic research}\ (\BPG~39-64).
\PrintBackRefs{\CurrentBib}

\bibitem [\protect \citeauthoryear {%
P\"ahtz%
, Kok%
, Parteli%
\BCBL {}\ \BBA {} Herrmann%
}{%
P\"ahtz%
\ \protect \BOthers {.}}{%
{\protect \APACyear {2013}}%
}]{%
Pahtz_1}
\APACinsertmetastar {%
Pahtz_1}%
\begin{APACrefauthors}%
P\"ahtz, T.%
, Kok, J\BPBI F.%
, Parteli, E\BPBI J\BPBI R.%
\BCBL {}\ \BBA {} Herrmann, H\BPBI J.%
\end{APACrefauthors}%
\unskip\
\newblock
\APACrefYearMonthDay{2013}{}{}.
\newblock
{\BBOQ}\APACrefatitle {Flux Saturation Length of Sediment Transport} {Flux
  saturation length of sediment transport}.{\BBCQ}
\newblock
\APACjournalVolNumPages{Phys. Rev. Lett.}{111}{}{218002}.
\PrintBackRefs{\CurrentBib}

\bibitem [\protect \citeauthoryear {%
Parteli%
, Andrade%
\BCBL {}\ \BBA {} Herrmann%
}{%
Parteli%
\ \protect \BOthers {.}}{%
{\protect \APACyear {2011}}%
}]{%
Parteli3}
\APACinsertmetastar {%
Parteli3}%
\begin{APACrefauthors}%
Parteli, E\BPBI J\BPBI R.%
, Andrade, J\BPBI S.%
\BCBL {}\ \BBA {} Herrmann, H\BPBI J.%
\end{APACrefauthors}%
\unskip\
\newblock
\APACrefYearMonthDay{2011}{Oct}{}.
\newblock
{\BBOQ}\APACrefatitle {Transverse Instability of Dunes} {Transverse instability
  of dunes}.{\BBCQ}
\newblock
\APACjournalVolNumPages{Phys. Rev. Lett.}{107}{}{188001}.
\newblock
\begin{APACrefURL}
  \url{https://link.aps.org/doi/10.1103/PhysRevLett.107.188001}
  \end{APACrefURL}
\newblock
\begin{APACrefDOI} \doi{10.1103/PhysRevLett.107.188001} \end{APACrefDOI}
\PrintBackRefs{\CurrentBib}

\bibitem [\protect \citeauthoryear {%
Parteli%
\ \protect \BOthers {.}}{%
Parteli%
\ \protect \BOthers {.}}{%
{\protect \APACyear {2014}}%
}]{%
Parteli4}
\APACinsertmetastar {%
Parteli4}%
\begin{APACrefauthors}%
Parteli, E\BPBI J\BPBI R.%
, Dur{\'a}n, O.%
, Bourke, M\BPBI C.%
, Tsoar, H.%
, P{\"o}schel, T.%
\BCBL {}\ \BBA {} Herrmann, H.%
\end{APACrefauthors}%
\unskip\
\newblock
\APACrefYearMonthDay{2014}{}{}.
\newblock
{\BBOQ}\APACrefatitle {Origins of barchan dune asymmetry: Insights from
  numerical simulations} {Origins of barchan dune asymmetry: Insights from
  numerical simulations}.{\BBCQ}
\newblock
\APACjournalVolNumPages{Aeol. Res.}{12}{}{121-133}.
\PrintBackRefs{\CurrentBib}

\bibitem [\protect \citeauthoryear {%
Parteli%
, Dur\'an%
\BCBL {}\ \BBA {} Herrmann%
}{%
Parteli%
\ \protect \BOthers {.}}{%
{\protect \APACyear {2007}}%
}]{%
Parteli}
\APACinsertmetastar {%
Parteli}%
\begin{APACrefauthors}%
Parteli, E\BPBI J\BPBI R.%
, Dur\'an, O.%
\BCBL {}\ \BBA {} Herrmann, H\BPBI J.%
\end{APACrefauthors}%
\unskip\
\newblock
\APACrefYearMonthDay{2007}{Jan}{}.
\newblock
{\BBOQ}\APACrefatitle {Minimal size of a barchan dune} {Minimal size of a
  barchan dune}.{\BBCQ}
\newblock
\APACjournalVolNumPages{Phys. Rev. E}{75}{}{011301}.
\newblock
\begin{APACrefURL} \url{https://link.aps.org/doi/10.1103/PhysRevE.75.011301}
  \end{APACrefURL}
\newblock
\begin{APACrefDOI} \doi{10.1103/PhysRevE.75.011301} \end{APACrefDOI}
\PrintBackRefs{\CurrentBib}

\bibitem [\protect \citeauthoryear {%
Parteli%
, Durán%
, Tsoar%
, Schwämmle%
\BCBL {}\ \BBA {} Herrmann%
}{%
Parteli%
\ \protect \BOthers {.}}{%
{\protect \APACyear {2009}}%
}]{%
Parteli6}
\APACinsertmetastar {%
Parteli6}%
\begin{APACrefauthors}%
Parteli, E\BPBI J\BPBI R.%
, Durán, O.%
, Tsoar, H.%
, Schwämmle, V.%
\BCBL {}\ \BBA {} Herrmann, H\BPBI J.%
\end{APACrefauthors}%
\unskip\
\newblock
\APACrefYearMonthDay{2009}{}{}.
\newblock
{\BBOQ}\APACrefatitle {Dune formation under bimodal winds} {Dune formation
  under bimodal winds}.{\BBCQ}
\newblock
\APACjournalVolNumPages{Proc. Natl. Acad. Sci. U.S.A.}{106}{52}{22085-22089}.
\PrintBackRefs{\CurrentBib}

\bibitem [\protect \citeauthoryear {%
Parteli%
\ \BBA {} Herrmann%
}{%
Parteli%
\ \BBA {} Herrmann%
}{%
{\protect \APACyear {2007}}%
}]{%
Parteli2}
\APACinsertmetastar {%
Parteli2}%
\begin{APACrefauthors}%
Parteli, E\BPBI J\BPBI R.%
\BCBT {}\ \BBA {} Herrmann, H\BPBI J.%
\end{APACrefauthors}%
\unskip\
\newblock
\APACrefYearMonthDay{2007}{Oct}{}.
\newblock
{\BBOQ}\APACrefatitle {Dune formation on the present Mars} {Dune formation on
  the present mars}.{\BBCQ}
\newblock
\APACjournalVolNumPages{Phys. Rev. E}{76}{}{041307}.
\newblock
\begin{APACrefURL} \url{https://link.aps.org/doi/10.1103/PhysRevE.76.041307}
  \end{APACrefURL}
\newblock
\begin{APACrefDOI} \doi{10.1103/PhysRevE.76.041307} \end{APACrefDOI}
\PrintBackRefs{\CurrentBib}

\bibitem [\protect \citeauthoryear {%
C.~Sauermann%
, Rognon%
, Poliakov%
\BCBL {}\ \BBA {} Herrmann%
}{%
C.~Sauermann%
\ \protect \BOthers {.}}{%
{\protect \APACyear {2000}}%
}]{%
Sauermann_1}
\APACinsertmetastar {%
Sauermann_1}%
\begin{APACrefauthors}%
Sauermann, C.%
, Rognon, P.%
, Poliakov, A.%
\BCBL {}\ \BBA {} Herrmann, H\BPBI J.%
\end{APACrefauthors}%
\unskip\
\newblock
\APACrefYearMonthDay{2000}{}{}.
\newblock
{\BBOQ}\APACrefatitle {The shape of the barchan dunes of {S}outhern {M}orocco}
  {The shape of the barchan dunes of {S}outhern {M}orocco}.{\BBCQ}
\newblock
\APACjournalVolNumPages{Geomorphology}{36}{}{47-62}.
\PrintBackRefs{\CurrentBib}

\bibitem [\protect \citeauthoryear {%
G.~Sauermann%
, Kroy%
\BCBL {}\ \BBA {} Herrmann%
}{%
G.~Sauermann%
\ \protect \BOthers {.}}{%
{\protect \APACyear {2001}}%
}]{%
Sauermann_4}
\APACinsertmetastar {%
Sauermann_4}%
\begin{APACrefauthors}%
Sauermann, G.%
, Kroy, K.%
\BCBL {}\ \BBA {} Herrmann, H\BPBI J.%
\end{APACrefauthors}%
\unskip\
\newblock
\APACrefYearMonthDay{2001}{Aug}{}.
\newblock
{\BBOQ}\APACrefatitle {Continuum saltation model for sand dunes} {Continuum
  saltation model for sand dunes}.{\BBCQ}
\newblock
\APACjournalVolNumPages{Phys. Rev. E}{64}{}{031305}.
\newblock
\begin{APACrefURL} \url{https://link.aps.org/doi/10.1103/PhysRevE.64.031305}
  \end{APACrefURL}
\newblock
\begin{APACrefDOI} \doi{10.1103/PhysRevE.64.031305} \end{APACrefDOI}
\PrintBackRefs{\CurrentBib}

\bibitem [\protect \citeauthoryear {%
Schw{\"a}mmle%
\ \BBA {} Herrmann%
}{%
Schw{\"a}mmle%
\ \BBA {} Herrmann%
}{%
{\protect \APACyear {2005}}%
}]{%
Schwammle}
\APACinsertmetastar {%
Schwammle}%
\begin{APACrefauthors}%
Schw{\"a}mmle, V.%
\BCBT {}\ \BBA {} Herrmann, H\BPBI J.%
\end{APACrefauthors}%
\unskip\
\newblock
\APACrefYearMonthDay{2005}{}{}.
\newblock
{\BBOQ}\APACrefatitle {A model of Barchan dunes including lateral shear stress}
  {A model of barchan dunes including lateral shear stress}.{\BBCQ}
\newblock
\APACjournalVolNumPages{Eur. Phys. J. E}{16}{1}{57-65}.
\PrintBackRefs{\CurrentBib}

\bibitem [\protect \citeauthoryear {%
Wenzel%
\ \BBA {} Franklin%
}{%
Wenzel%
\ \BBA {} Franklin%
}{%
{\protect \APACyear {2019}}%
}]{%
Wenzel}
\APACinsertmetastar {%
Wenzel}%
\begin{APACrefauthors}%
Wenzel, J\BPBI L.%
\BCBT {}\ \BBA {} Franklin, E\BPBI M.%
\end{APACrefauthors}%
\unskip\
\newblock
\APACrefYearMonthDay{2019}{}{}.
\newblock
{\BBOQ}\APACrefatitle {Velocity fields and particle trajectories for bed load
  over subaqueous barchan dunes} {Velocity fields and particle trajectories for
  bed load over subaqueous barchan dunes}.{\BBCQ}
\newblock
\APACjournalVolNumPages{Granular Matter}{21}{}{321-334}.
\PrintBackRefs{\CurrentBib}

\bibitem [\protect \citeauthoryear {%
Zhang%
, Yang%
, Rozier%
\BCBL {}\ \BBA {} Narteau%
}{%
Zhang%
\ \protect \BOthers {.}}{%
{\protect \APACyear {2014}}%
}]{%
Zhang_D}
\APACinsertmetastar {%
Zhang_D}%
\begin{APACrefauthors}%
Zhang, D.%
, Yang, X.%
, Rozier, O.%
\BCBL {}\ \BBA {} Narteau, C.%
\end{APACrefauthors}%
\unskip\
\newblock
\APACrefYearMonthDay{2014}{}{}.
\newblock
{\BBOQ}\APACrefatitle {Mean sediment residence time in barchan dunes} {Mean
  sediment residence time in barchan dunes}.{\BBCQ}
\newblock
\APACjournalVolNumPages{J. Geophys. Res.: Earth Surf.}{119}{3}{451--463}.
\PrintBackRefs{\CurrentBib}

\end{thebibliography}

\end{document}